\documentclass[a4paper,12pt]{article}
\pdfoutput=1

\usepackage{jheppub}
\usepackage{xcolor}

\usepackage[T1]{fontenc} 
\usepackage{amscd}
\usepackage[all,cmtip]{xy}
\usepackage{mathrsfs}

\usepackage{subfigure}
\usepackage{dsdshorthand}
\usepackage{simplewick}
\usepackage{booktabs,multirow}

\newcommand*\DAlambert{\mathop{}\!\mathbin\Box}

\title{\boldmath A large-$N$ tensor model with four supercharges}
\author[a]{Davide Lettera}
\author[a]{Alessandro Vichi}
\affiliation[a]{\it Department of Physics, University of Pisa, \\ Largo Bruno Pontercorvo 3, 56127 Pisa, Italy}

\abstract{We study a supersymmetric tensor model with four supercharges and  $O(N)^3$ global symmetry. The model is based on a chiral scalar superfield with three indices and quartic tetrahedral interaction in the superpotential, which is relevant below three dimensions. In the large-$N$ limit the model is dominated by melonic diagrams. We solve the Dyson-Schwinger equations in superspace for generic $d$ and extract the dimension of the chiral field and the dimensions of bilinear operators transforming in various representations of $O(N)^3$. We find that all operator dimensions are real and above the unitarity bound for  $1<d<3$. Our results also agree with perturbative results in  $3-\varepsilon$ expansion. Finally, we extract the large spin behaviour of  bilinear operators and discuss the connection with lightcone bootstrap.}

\begin{document}
\maketitle
\flushbottom

\section{Introduction}
\begin{figure}[htb!]
	\centering
	\includegraphics[width=0.4\linewidth]{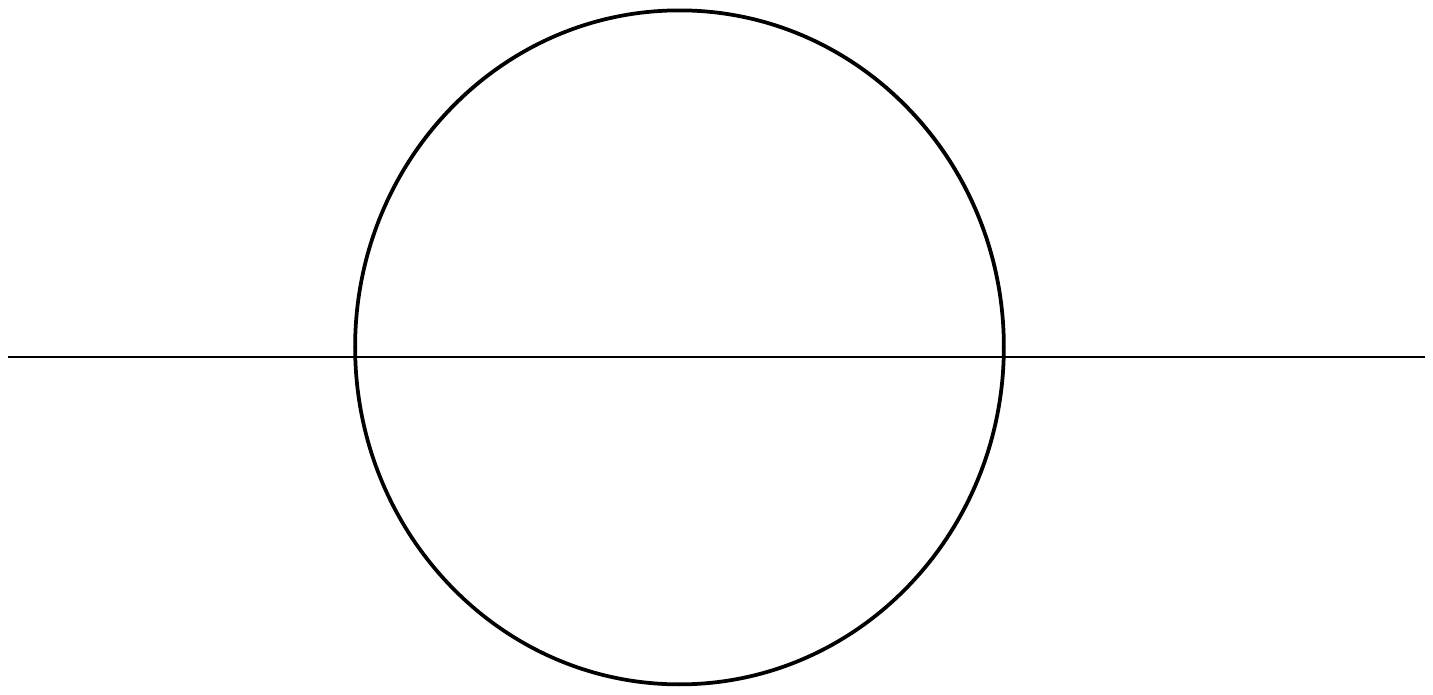}
	\caption{Melonic diagram contributing to the 2-point function in a tensor model theory with three indexes and quartic interaction.}
	\label{fig:fundamentalmelon}
\end{figure}

Quantum field theories with a large number of components $N$ are extremely fascinating objects: despite they often are strongly-interacting systems, in certain cases it is possible to find exact solutions.
Indeed, radiative corrections are weighted by different powers of $N$, depending on their topology. Thus, in the large $N$ limit, only a subset of all possible diagrams dominates and sometimes it is possible to resum them completely.\\
The most famous example of this mechanism is perhaps the $O(N)$ vector model \cite{ZinnJustin1,Moshe:2003xn}, a theory of $N$ scalar fields $\phi_a$ with quartic interaction $g \left( \phi_a\phi_a \right)^2$. The dominant diagrams are called snail diagrams and the expansion parameter becomes $gN$; therefore, the large $N$ theory is defined by keeping $g N$ fixed. The theory is exactly solvable with various techniques.\\ \indent
On the other hand, there are other examples where the dominant Feynman diagrams are still too many and too different, and is not possible to exactly sum them. A notable example are matrix models with $N^2$ interacting scalar fields. In this case, the large $N$ limit is dominated by planar diagrams, with expansion parameter $g N^2$ \cite{GT1}.\\ \indent
In recent years, there has been an increasing interest in a new large-$N$ behaviour: the melonic limit of tensor models \cite{KlebPop,Gurau:2019qag,BenedettiMelonicCFT,Klebanov:2016xxf}. 
This particular limit arises when we study models with $N^r$ fields transforming in the fundamental representation of $O(N)^r$. Regardless of the exact structure of the theory \cite{Gurau:2009tw,Witten1,Bonzom:2012hw,GiombiKleb,GurauBen2,Benedetti:2019rja,Giombi:2018qgp}, a proper choice of the interaction vertex selects a well defined subset of diagrams dominating in the large $N$ limit: the melonic diagrams, see for instance Figure~\ref{fig:fundamentalmelon}.

Melonic diagrams are precisely the subset of planar diagrams that dominates the large N limit of the Sachdev-Ye-Kitaev (SYK) model \cite{MaldacenaStanford,ROSsyk,GrossRoss,POL}.
Remarkably, sometimes it is possible to sum them exactly by means of self-consistency relations: the Dyson-Schwinger equation (DSE) for the 2-point (2pt) function and for the 3-point (3pt) function.

In this paper we study a supersymmetric theory of chiral and anti-chiral fields, transforming in the fundamental representation of the global symmetry group $\mathcal{G}=O(N)^3$. The index structure of this theory is identical to that of a bosonic model in a non-supersymmetric theory. As a consequence, the formal proof of melonic dominance still holds when applied to super-diagrams.\\
In section~\ref{N=1} we provide a partial review of a similar model with two supercharges,  which has been first studied in \cite{POP1}. We use this model as a warm up exercise. In section~\ref{N=2} we generalize  the model to $4$ supercharges, equivalent to $\mathcal N=2$ supersymmetry in three dimensions. This model was first mentioned in \cite{Klebanov:2016xxf} and then studied further in \cite{POP1}. We compute the dimension of the chiral superfield $\Delta_\Phi$  and investigate the spectrum of all bilinear operators (scalar and spin-$\ell$), transforming in the $10$ possible irreducible representation  (irreps) of $(v,v,v) \otimes (v,v,v)$, where $v$ is the fundamental representation of $O(N)$. The technical details of the calculations are very similar to those of section~\ref{N=1}, but  the extended supersymmetry gives us a better control on the results. As an example, we checked the existence of conserved multiplets associated to the stress tensor and the global symmetry current, with the correct spin and irrep. Moreover, we compared with existing results for $3-\varepsilon$ dimensions and find perfect agreement. In this regime we are also able to check multiplet recombination phenomena.

Finally, we explored the large spin behaviour of bilinear operators and compared with the expected behaviour of double-trace operators in a conformal field theory (CFT) \cite{Komargodski:2012ek,Fitzpatrick:2012yx,Alday:2016njk,Alday:2016jfr,Caron-Huot:2017vep}. We find that the Regge trajectories of three irreps have dimension  $\Delta=2\Delta_\Phi +\ell +2n +\gamma(\ell)$, with $\gamma(\ell)$ is a non vanishing function of the spin $\ell$, even in the infinite $N$ limit. It is worth noticing that this behaviour is different from what observed in vector models, where all double trace operators have dimension exactly equal to $ \ell+d-2$ at infinite $N$.\\
 In this respect, tensor models offer an interesting opportunity to study non-trivial, exact, Regge trajectories.
\\

\section{Warm up: $\mathcal{N}=1$ tetrahedral model} \label{N=1}
The simplest possibility of a supersymmetric tetrahedral model is with $\mathcal{N}=1$ \cite{POP1}. The fundamental field is a real scalar $\Phi_{a b c}(x,\theta)$. In analogy with the tetrahedral model \cite{GiombiKleb,GurauBen2}, each global symmetry index ($a$, $b$ and $c$) transforms in the fundamental representation of $O(N)$. The model is defined by the action \cite{POP1},
\begin{equation} \label{SactionN1}
S[\Phi]=\int [d^d x][d ^2\theta] (- \frac{1}{2}  \Phi_{a b c} D^2 \Phi_{a b c} +  \frac{1}{4} g \Phi_{a b c} \Phi_{a d e} \Phi_{f b e} \Phi_{f d c} ).
\end{equation}
where $[d^2 \theta]$ is the invariant measure on superspace and $D$ is a covariant derivative. After integrating over the Grassmann variables, we get an action in ordinary space-time with a $\phi^6$ interaction (and $\phi^2\bar\psi\psi$ terms). This means that the interaction is exactly marginal in $d=3$ and irrelevant in $d=4$.\\
The details of 3d $\mathcal{N}=1$ superspace are discussed in appendix \ref{appendixA}.

\subsection{DSE for $\mathcal{N}=1$}
The Dyson-Schwinger equation in the large $N$ limit can be written exactly in the same way as in the bosonic tetrahedral model case \cite{GiombiKleb,KlebPop,Klebanov:2016xxf,BenedettiMelonicCFT}. The propagator and exact 2pt function can be written as
\begin{align}
		&\langle \Phi_{abc}(-p,\theta_1) \Phi_{a'b'c'}(p,\theta_2) \rangle_0= G_0(p,\theta_1,\theta_2)\delta_{a a'} \delta_{b b'} \delta_{c c'},\\	
		&\langle \Phi_{abc}(-p,\theta_1) \Phi_{a'b'c'}(p,\theta_2) \rangle= G(p,\theta_1,\theta_2)\delta_{a a'} \delta_{b b'} \delta_{c c'},
\end{align} 
\\
while the DSE reads (schematically)
\begin{equation} \label{SchematicDSE}
G=G_0+\lambda^2G_{0} G^3 G.
\end{equation}
\begin{figure}[tb!]
	\centering
	\includegraphics[width=0.9\linewidth]{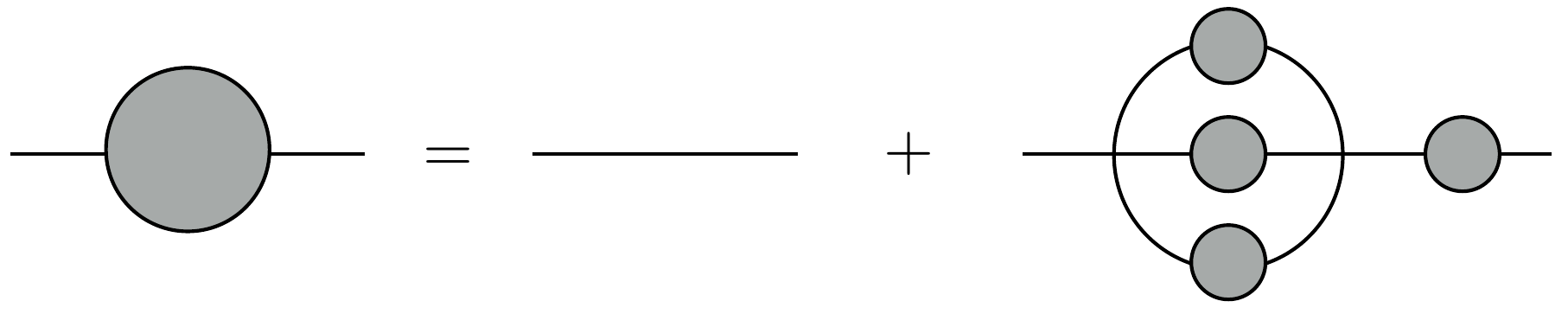}
	\caption{Diagrammatic representation of the DSE. The line with the full ball represents the two-point function while the line alone represents the propagator.}
	\label{fig:dse}
\end{figure}

In equation (\ref{SchematicDSE}) $\lambda^2=N^3 g^2  $ is kept fixed while $N$ approaches infinity.
The only difference from the bosonic model is in the integration over super space-time.
In appendix \ref{appendixB1} we review the computation of $G_0({p,\theta_1,\theta_2})$ in details and we find
\begin{equation} \label{propagator}
	G_0(p,\theta_1,\theta_2)=\frac{D^2\delta^{2} ( \theta_1 - \theta_2 )}{p^2}=\frac{e^{-\theta_1\gamma^\mu \theta_2 p_\mu} }{p^2}.
\end{equation}
The form of two point functions are constrained by superconformal symmetry\footnote{In coordinate space \cite{POP1}: $G(x,\theta_1,\theta_2)=\frac{B}{|x^\mu-(\theta_2 \gamma^\mu \theta_1)|^{2\Delta_\Phi}}$, where $x^\mu = x_2^\mu-x_1^\mu$.} and are given by
\\
\begin{equation} \label{N12PT}
G(p,\theta_1,\theta_2)=A\frac{D^2\delta^{2} ( \theta_1 - \theta_2 )}{p^{2\Delta}} .
\end{equation}
\\
Substituting \eqref{propagator} and \eqref{N12PT}  in equation~\eqref{SchematicDSE} it is possible to find the values of  $\Delta$ that solve the DSE:

\begin{align} \nonumber 
&A\frac{D^2\delta^2(\theta_1-\theta_2)}{p^{2\Delta}}=\frac{D^2\delta^2(\theta_1-\theta_2)}{p^2}+A^4 \lambda^2 \int d^2 \theta' d^2 \theta'' \int \frac{d^d q}{(2\pi)^d} \frac{d^d k}{(2\pi)^d}
\\
&\frac{D^2\delta^2(\theta_1-\theta')}{p^2} \frac{D^2\delta^2(\theta'-\theta'')}{|p-q-k|^{2\Delta}} \frac{D^2\delta^2(\theta'-\theta'')}{q^{2\Delta}}
\frac{D^2\delta^2(\theta'-\theta'')}{k^{2\Delta}}
\frac{D^2\delta^2(\theta''-\theta_2)}{p^{2\Delta}}. \label{DSEN1}
\end{align}
\\
As it is standard in literature \cite{ROSsyk,GrossRoss,POL,MaldacenaStanford}, the (l.h.s.) of (\ref{DSEN1}) can be neglected in the IR limit provided that $\Delta<1$
\\
\begin{equation} \label{DSEIRN1}
-\frac{D^2\delta^2(\theta_2-\theta_1)}{p^2}=A^4 \lambda^2  \int \frac{d^d q}{(2\pi)^d} \frac{d^d k}{(2\pi)^d}
\frac{\mathcal{J}(p,q,k,\theta_1,\theta_2)}{p^2|p-k_1-k_2|^{2\Delta}k_1^{2\Delta}
k_2^{2\Delta}p^{2\Delta}},
\end{equation}
\\
where $\mathcal{J}(p,q,k,\theta_1,\theta_2)$ takes into account the integration of Grassmann variables and we can write it schematically as\footnote{In this form the dependence on the momenta is hidden and the reader should remember that the covariant derivative $D$ depends on the momentum.
We can reconstruct the precise moment dependence of each $D^2\delta^2(\theta_a-\theta_b)$ factor by looking at equation (\ref{DSEN1}). }
\\
\begin{equation} \label{JschematicDef}
\mathcal{J}(p,q,k,\theta_1,\theta_2)= \int d^2\theta' d^2 \theta'' D^2\delta^2(\theta_1-\theta') \left(D^2\delta^2(\theta'-\theta'')\right)^3 D^2\delta^2(\theta''-\theta_2).
\end{equation}
\\
Using equation (\ref{propagator}) it is easy to see that $k$ and $q$ simplify, this means that $\mathcal{J}(p,q,k,\theta_1,\theta_2)$ depends on $p$, $\theta_1$ and $\theta_2$ only. Then, by dimensional analysis, the result is constrained to be of the form
\\
\begin{equation}
\mathcal{J}(p,q,k,\theta_1,\theta_2)\propto p^2 D^2 \delta^2(\theta_2-\theta_1)
\end{equation}
\\
and this is enough to find the parameter $\Delta$ in equation~\eqref{DSEIRN1}. This method however does not fix the overall factor. Hence we will compute $\mathcal{J}(p,k_1,k_2,\theta_1,\theta_2)$ performing explicitly the integration in appendix \ref{appendixB2}. We find
\\
\begin{equation}
	\mathcal{J}(p,q,k,\theta_1,\theta_2)=-p^2 D^2 \delta^2(\theta_2-\theta_1),
\end{equation}
\\
that is the expected results with the correct proportionality constant.
Substituting it into the DSE we find
\\
\begin{align}
&\Delta=\frac{d+1}{4},\qquad\qquad A^4=\frac{(4 \pi)^d}{\lambda^2}\frac{\Gamma^3( \frac{d+1}{4} ) \Gamma ( 3 \frac{d-1}{4} )}{\Gamma^3 ( \frac{d-1}{4} ) \Gamma( \frac{3-d}{4} )}.
\end{align}
\\
The parameter $\Delta$ is easily related to the scaling dimension of $\Phi$ by
\begin{equation}
	\Delta_{\Phi}=\frac{d}{2}-\Delta=\frac{d-1}{4}\,.
\end{equation}
Finally, we can compute the anomalous dimension in the $3-\varepsilon$ expansion. Since the dimension of the free field would be $ \Delta_{0}=\frac{d-2}{2} $, we can write $\Delta_\Phi=\Delta_0+\gamma_\Phi$ and  get 
\begin{equation}
\gamma_\Phi=\frac{\varepsilon}{4}\,.
\end{equation}

\begin{figure}[t]
	\centering
	\includegraphics[width=0.8\textwidth]{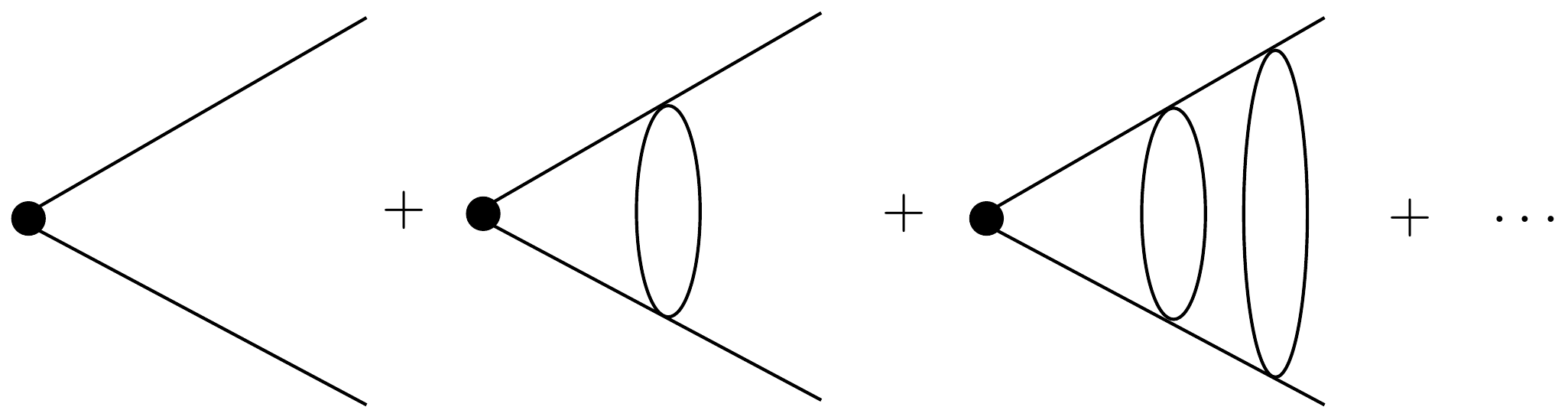}
	\caption{Radiative contributions to the three point function in the melonic limit.}
	\label{fig:radiativecontributions3pt}
\end{figure}
\subsection{Bilinear operators}
There are only two possible kinds of spin-$0$ singlet bilinear operators:
\\
\begin{align}
& {\mathcal O}^h_{1}= \Phi_{a b c}\DAlambert^h  \Phi_{a b c},\\
& {\mathcal O}^h_{2}= \Phi_{a b c}\DAlambert^h D^2 \Phi_{a b c}.
\end{align}
\\
The insertion of others $D^2$ changes an $\mathcal O_{2}$  type operator into an $\mathcal O_{1}$ (and viceversa) type with $h$ increased by 1 and do not leads to different operators.\\
The form of three point functions is constrained by the superconformal symmetry \cite{PARK1,ATA1,NIZAMI}: following \cite{POP1} we use the ansatz
\begin{align}\label{GBBansatz}
&V_{1}(p,\theta,\theta')=\langle {\mathcal O}_{1}(0) \Phi(-p,\theta) \Phi(p,\theta') \rangle=\frac{D^2 \delta^2(\theta-\theta')}{p^{2\Delta+\Delta_{\mathcal O}}},\\
&V_{2}(-p,\theta,\theta')=\langle {\mathcal O}_{2}(0) \Phi(-p,\theta) \Phi(p,\theta') \rangle=\frac{\delta^2(\theta-\theta')}{p^{2\Delta+(\Delta_{\mathcal O}-1)}}.\label{GFFansatz}
\end{align}
One can easily convince himself that the ansatz  (\ref{GBBansatz}) and (\ref{GFFansatz}) are correct by computing  the above 3pt functions in free theory.
Notice that in equations (\ref{GBBansatz}) and (\ref{GFFansatz}) we have set the momentum and the Grassmann variables of $\mathcal{O}_{1,2}$ to zero. In the direct space this is equivalent to set the space coordinate of $\mathcal{O}_{1,2}$ to infinity, as it is standard in literature \cite{GrossRoss,POL,ROSsyk}.

The exact 3pt function in the large-$N$ limit is dominated by an infinite sum of diagrams as shown in Figure~\ref{fig:radiativecontributions3pt}. This structure is exactly the same as found in bosonic tensor models: the only difference is in the precise form of 2pt functions and 3pt functions, and the fact that integrations are intended in superspace. 

\begin{figure}[t!]
	\centering
	\includegraphics[width=0.2\textwidth]{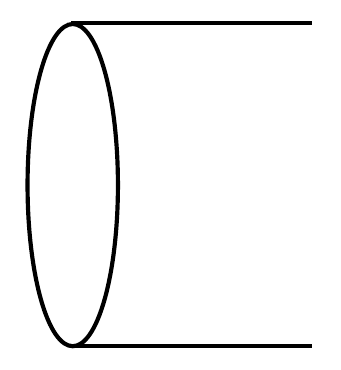}
	\caption{Kernel operator.}
	\label{fig:kerneloperator}
\end{figure}

\begin{figure}[htb!] 
	\centering
	\includegraphics[width=0.3\linewidth]{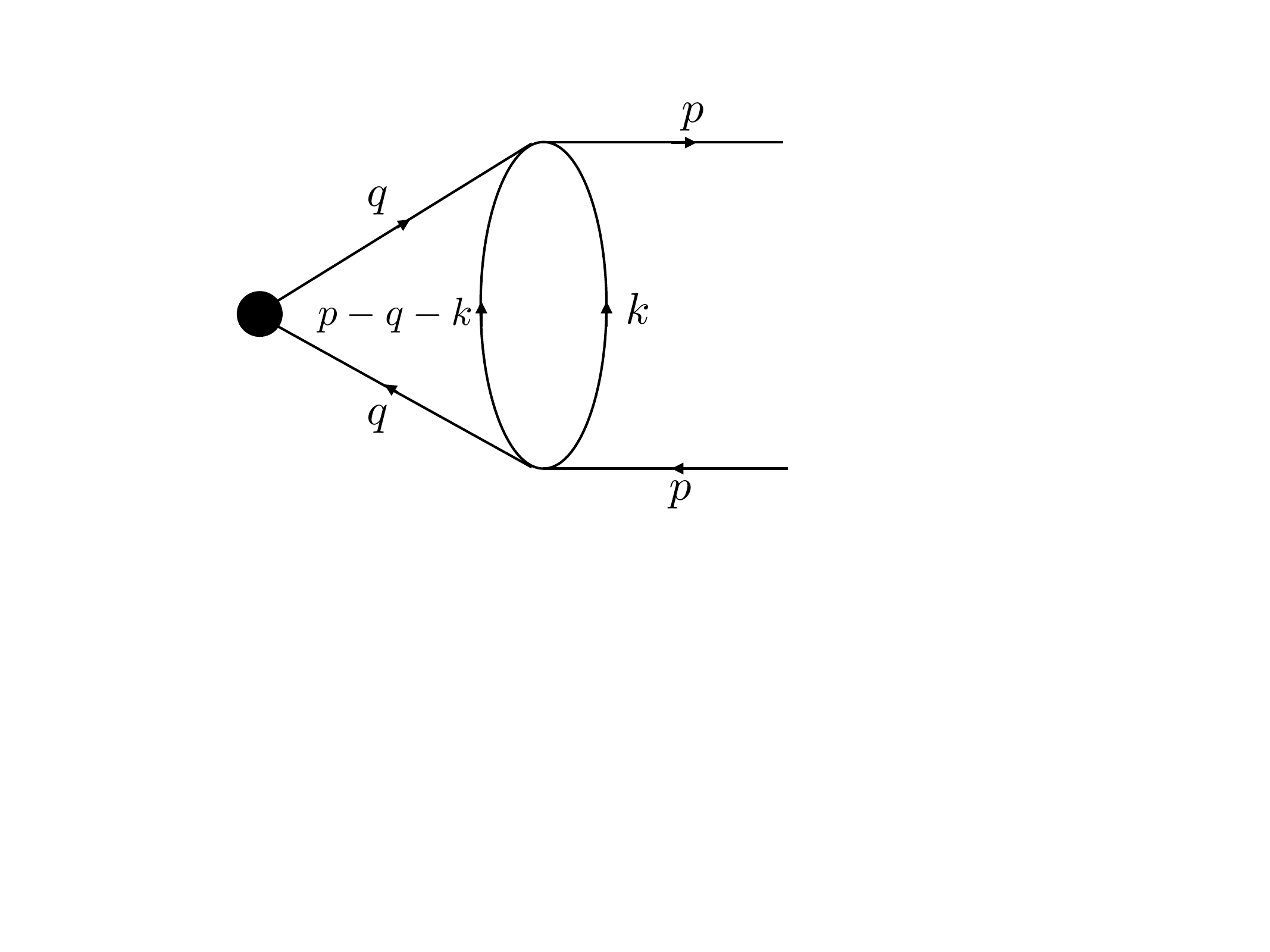}
	\caption{Diagrammatic representation of equation (\ref{MomentaKernelN1}).}
	\label{fig:appliedkernelmomenta}
\end{figure}

\begin{figure}[tb!]
	\centering
	\includegraphics[width=0.6\linewidth]{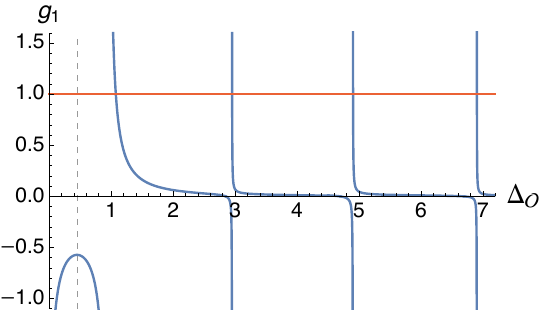}
	\caption{Solution of eigenvalue problem  $g_{1}(\Delta_{\mathcal O})=1$ in $d=2.9$ dimensions. The vertical dashed line corresponds to the unitarity bound.}
	\label{plotgff}
\end{figure}

Since the 3pt function is the infinite sum of all ladder diagrams, it must be an eigenfunction with eigenvalue $1$ of the operator  $\mathcal{K}$ that ``adds a rung to the ladder''.
In the coordinates space the Kernel operator $\mathcal{K}$ reads
\\
\begin{equation}
	\mathcal{K}(x_1\theta_1,x_2 \theta_2,x_3, \theta_3,x_4, \theta_4)=3 \lambda^2 G(x_{13},\theta_1,\theta_3) G(x_{24},\theta_2,\theta_4)G(x_{34},\theta_3,\theta_4)^2
\end{equation}
\\
and its diagrammatic representation is in Figure~\ref{fig:kerneloperator}.
\\
The eigenvalue equation written in coordinates space takes the following form 
\\
\begin{align} \label{EigenvalueEq}
\mathcal{K}V_i(x_1\theta_1,x_2\theta_2)& \equiv \int [d^d x' d^2\theta'] [d^dx'' d^2 \theta''] \mathcal{K}(x_1\theta_1,x_2\theta_2,x'\theta',x''\theta'')V_{i}(x'\theta',x''\theta'')\nonumber \\
&=g_{i}(\Delta_{\mathcal O}) V_{i}(x_1\theta_1,x_2\theta_2).
\end{align}
\\
We start solving it for for $V_{1}$ and we find easier working in the momentum space. Then equation (\ref{EigenvalueEq}) becomes
\\
\begin{align} \nonumber 
&\mathcal{K}V_1(p,\theta_1,\theta_2)=3 \lambda^2 A^4  \int d^2 \theta' d^2 \theta'' \int \frac{d^d q}{(2\pi)^d} \frac{d^d k}{(2\pi)^d}
\\
&\times \frac{D^2\delta^2(\theta_1-\theta')}{p^{2\Delta}} \frac{D^2\delta^2(\theta''-\theta_2)}{p^{2\Delta}} \frac{D^2\delta^2(\theta'-\theta'')}{q^{2\Delta}}
\frac{D^2\delta^2(\theta'-\theta'')}{|p-q-k|^{2\Delta}}
\frac{D^2\delta^2(\theta'-\theta'')}{k^{2\Delta+\Delta_{\mathcal O}}}\label{MomentaKernelN1}
\end{align}
\\
and its diagrammatic representation is in Figure~\ref{fig:appliedkernelmomenta}.
\\
\\
The contribution of the superspace integration is exactly equal to $\mathcal{J}(p,q,k,\theta_1,\theta_2)$ and one gets
\begin{equation} \label{N1KernelMom}
	\mathcal{K}V_{1}(p, \theta_1, \theta_2)=-3\lambda^2 A^4 \frac{D^2 \delta^2 (\theta_2-\theta_1)}{p^{4\Delta-2}} \int \frac{d^d q}{(2\pi)^d}\frac{d^d k}{(2\pi)^d}\frac{1}{q^{2\Delta}k^{2\Delta+\Delta_{\mathcal O}}|p-q-k|^{2\Delta}}.
\end{equation}
It is now necessary to use the value $\Delta=\frac{d+1}{4}$ and the following known integral 
\begin{align} \label{MomentumKnownInt}
&\int \frac{d^d k}{(2\pi)^d} \frac{1}{k^{2\alpha}(k+p)^{2\beta}}=\frac{L_d(\alpha,\beta)}{(2\pi)^d} \frac{1}{(p^2)^{\alpha+\beta-\frac{d}{2}}} ,\\
&L_d(\alpha_1,\alpha_2)=\pi^{\frac{d}{2}} \frac{\Gamma(\frac{d}{2}-\alpha_1) \Gamma(\frac{d}{2}-\alpha_2) \Gamma(\frac{d}{2}-\alpha_3)}{\Gamma(\alpha_1) \Gamma(\alpha_2) \Gamma(\alpha_3)}, \ \ \ \ \ \ \ \alpha_1+\alpha_1+\alpha_3=d. \label{MomentaIntegral}
\end{align}
\\
We finally get
\begin{align}
&\mathcal{K}V_{1}(p, \theta_1, \theta_2)=g_{1}(\Delta_{\mathcal O})\frac{D^2\delta^2(\theta_2-\theta_1)}{p^{2\Delta+\Delta_{\mathcal O}}},\\
&g_{1}(\Delta_{\mathcal O})=-3\frac{\Gamma(\frac{d+1}{4})\Gamma(3\frac{d-1}{4})\Gamma(\frac{d-1}{4}-\frac{\Delta_{\mathcal O}}{2})\Gamma(\frac{3-d}{4}-\frac{\Delta_{\mathcal O}}{2})}{\Gamma(\frac{d-1}{4})\Gamma(\frac{3-d}{4})\Gamma(\frac{d+1}{4}+\frac{\Delta_{\mathcal O}}{2}){\Gamma(3\frac{d-1}{4}-\frac{\Delta_{\mathcal O}}{2})}}.\label{GFF}
\end{align}
\\
The constrain $g_{1}(\Delta_{\mathcal O})=1$ fixes the possible scaling dimensions of primary operator and it can be solved only numerically. Looking at the plot in $d=2.9$ , see Figure~\ref{plotgff},
we clearly see that there are solutions around odd integers. This is reasonable, since these represents the dimensions of operators of the form $\sim \Phi \DAlambert^h \Phi$. \\
It is also possible to obtain the solution in $d=3-\varepsilon$ and we find that the lowest solution is 
\\
\begin{equation}\label{DeltaPhiPhi}
	\Delta_{\Phi \Phi}=1+\varepsilon + 3 \varepsilon^2+O(\varepsilon^3).
\end{equation}
\\

Next, we consider the solution of equation (\ref{EigenvalueEq}) for $V_2$. The only difference with respect to the previous case is in the superspace part of $V_{2}$
\begin{align} \nonumber 
	&\mathcal{K}V_{2}(p, \theta_1, \theta_2)=3 \lambda^2 A^4  \int d^2 \theta' d^2 \theta'' \int \frac{d^d q}{(2\pi)^d} \frac{d^d k}{(2\pi)^d}
	\\
	&\frac{D^2\delta^2(\theta_1-\theta')}{p^{2\Delta}} \frac{D^2\delta^2(\theta''-\theta_2)}{p^{2\Delta}} \frac{D^2\delta^2(\theta'-\theta'')}{q^{2\Delta}}
	\frac{D^2\delta^2(\theta'-\theta'')}{|p-q-k|^{2\Delta}}
	\frac{\delta^2(\theta'-\theta'')}{k^{2\Delta+(\Delta_{\mathcal O}-1)}}. \label{KV2}
\end{align}
\\
In this case the contribution of the integral over Grassmann variables differs from $\mathcal{J}(p,\theta_1,\theta_2)$
\\
\begin{align}\label{KBB}
&\mathcal{K}V_{2}(p, \theta_1 , \theta_2)=3\lambda^2 A^4 \frac{1}{p^{4\Delta}} \int \frac{d^dq}{(2\pi)^d}\frac{d^d k}{(2\pi)^d}\frac{\mathcal{J}_{2}(p,q,k,\theta_1,\theta_2)}{q^{2\Delta}k^{2\Delta+(\Delta_{\mathcal O}-1)}|p-q-k|^{2\Delta}},\\ \label{J2schematic}
&\mathcal{J}_{2}(p,q,k,\theta_1,\theta_2)= \int d^2\theta' d^2 \theta'' D^2\delta^2(\theta_1-\theta') \left( \left(D^2\delta^2(\theta'-\theta'')\right)^2 \delta(\theta'-\theta'') \right) D^2\delta^2(\theta''-\theta_2).
\end{align}
\\
The computation of $\mathcal{J}_{2}(p,q,k,\theta_1,\theta_2)$ can be found in appendix \ref{appendixB3} and the solution turns out to be
\begin{equation} \label{J2}
	\mathcal{J}_{2}(p,q,k,\theta_1,\theta_2)=-p^2\delta^2(\theta_2-\theta_1).
\end{equation}
Substituting equation (\ref{J2})  into (\ref{KBB}) we get
\\
\begin{equation}
\mathcal{K}V_{2}(p, \theta_1, \theta_2)=-3\lambda^2 A^4 \frac{\delta^2(\theta_2-\theta_1)}{p^{4\Delta}} \int \frac{d^dq}{(2\pi)^d}\frac{d^d k}{(2\pi)^d}\frac{1}{q^{2\Delta}k^{2\Delta+(\Delta_{\mathcal O}-1)}|p-q-k|^{2\Delta}}.
\end{equation}
\\
The above expression has exactly the same form of the eigenvalue equation of $V_{1}$, but with the substitution $\Delta_{\mathcal O} \rightarrow \Delta_{\mathcal O}-1$. Thanks to this remark we can skip other computations and state that correct constraint is
\\
\begin{equation}
	g_{2}(\Delta_{\mathcal O})=g_{1}(\Delta_{\mathcal O}-1)=1,
\end{equation}
\\
which have the same solution of $g_{1}(\Delta_{\mathcal O})=1$ but translated by one. In this case operator dimensions are close to even integer for  $d\sim 3$.\\
The lowest scaling dimension in $d=3-\varepsilon$  is
\\
\begin{equation}
\Delta_{\Phi D^2 \Phi}=2+\varepsilon+ 3\varepsilon^2+O(\varepsilon^3).
\end{equation}


\section{Tetrahedral model with $\mathcal{N}=2$ supersymmetry} \label{N=2}

Let us now move to the case of four supercharges, which would correspond to $\mathcal{N}=2$ in three dimensions. The $3d$ $\mathcal{N}=2$ super-algebra can be obtained from the 4d $\mathcal{N}=1$ one. Indeed, if we start from a theory with two real bi-spinors $Q_1$ and $Q_2$ (4 supercharges), the change of variables $Q=Q_1+iQ_2$ and $\bar{Q}=Q_1-iQ_2$ brings its algebra exactly in the same form of the $\mathcal{N}=1$ algebra in 4d (see appendix \ref{appendixA2}). This algebra has been studied in general $d$ between two and four dimensions in \cite{Bobev}.
\\The fundamental fields of the theory are a chiral field $\Phi_{abc}$ and an anti-chiral $\bar{\Phi}_{abc}$ with action
\begin{align} \label{SactionN2} 
&S[\Phi,\bar{\Phi}]=\int d^dx\int d^2\theta d^2\bar{\theta} \left(   \bar{\Phi}_{a b c} \Phi_{a b c}  \right)+\int d^dy d^2\theta W[\Phi] +\int d^d\bar{y}  d^2 \bar{\theta} W[\bar{\Phi}],\\ \nonumber
& W[\Phi]=\frac{1}{4} g \Phi_{a b c} \Phi_{a d e} \Phi_{f b e} \Phi_{f d c},
\end{align}
Since the combinatorial properties do not change from $\mathcal{N}=1$ to $\mathcal{N}=2$, we expect similar DSEs. On the other hand, this more (super-)symmetric model must satisfies more constraints and the integration over superspace changes. Thus, we  do expect few differences. 
\\
The covariant derivatives are defined in equations (\ref{DbarD}) and (\ref{DbarD2}) and satisfy the relation 
\\
\begin{equation}\label{AntiCommDbarD}
\{ D_{\alpha},\bar{D}_{\beta} \} \ = \ 2i \gamma^{\mu}_{\alpha \beta} \partial_{\mu}.
\end{equation}
\\
Exploiting the analogy with the four dimensional case we can write the propagator in the form
\\
\begin{align}
	\langle \Phi_{abc}(x_1,\theta_1,\bar{\theta}_1) \bar{\Phi}_{a'b'c'}(x_2,\theta_2,\bar{\theta}_2)\rangle_0&=\langle \Phi_{abc}(y_1,\theta_1)\bar{\Phi}_{a'b'c'}(\bar{y}_2,\bar{\theta}_2)\rangle_0 \nonumber\\
	&=\frac{e^{+i2(\bar{\theta}_2 \gamma^\mu \theta_1)\partial_\mu}}{-\DAlambert}\delta^3(\bar{y}_2-y_1)\delta_{aa'}\delta_{bb'}\delta_{cc'}.
\end{align} 
\\
While in momentum space it reads
\\
\begin{equation}
	\langle \Phi_{abc}(-p,\theta_1)\bar{\Phi}_{a'b'c'}(p,\bar{\theta}_2)\rangle_0=\frac{e^{-2(\bar{\theta}_2 \gamma^\mu \theta_1)p_\mu}}{p^2}\delta_{aa'}\delta_{bb'}\delta_{cc'}.
\end{equation}
\subsection{DSE for $\mathcal{N}=2$}
The DSE can be obtained following the same steps as in the $\mathcal{N}=1$ model.  We define the propagator $G_0$ and the two point function $G$ as
\\
\begin{align}
	&\langle \Phi_{abc}(-p,\theta_1) \bar{\Phi}_{a'b'c'}(p,\bar{\theta}_2) \rangle_0 \equiv G_0(p,\theta_1,\bar{\theta}_2)\delta_{a a'} \delta_{b b'} \delta_{c c'},\\	
	&\langle \Phi_{abc}(-p,\theta_1) \bar{\Phi}_{a'b'c'}(p,\bar{\theta}_2) \rangle \equiv G(p,\theta_1,\bar{\theta}_2)\delta_{a a'} \delta_{b b'} \delta_{c c'}.
\end{align} 
Since $\langle \Phi \Phi \rangle=\langle \bar{\Phi} \bar{\Phi}\rangle=0$, the only way to have a non-vanishing contribution to a melon diagram is to consider the insertion of one (and only one) $W(\Phi)$ and one (and only one) $W(\bar{\Phi})$. \\
We make the usual conformal ansatz for the two point function
\begin{equation}
G(p,\theta_1,\bar{\theta}_2)=\langle \Phi(-p,\theta_1)\bar{\Phi}(p,\bar{\theta}_2)\rangle=A_{2}\frac{e^{-2(\bar{\theta}_2 \gamma^\mu \theta_1)p_\mu}}{p^{2\Delta}},
\end{equation}
\\
and the resulting DSE is
\begin{align} \nonumber 
&A_2\frac{e^{2(\bar{\theta}_1 \gamma^\mu \theta_2)p_\mu}}{p^{2\Delta}}=\frac{e^{2(\bar{\theta}_1 \gamma^\mu \theta_2)p_\mu}}{p^2}+2A_2^4 \lambda^2 \int d^2 \theta' d^2 \bar{\theta}'' \int \frac{d^d q}{(2\pi)^d} \frac{d^d k}{(2\pi)^d}
\\
&\frac{e^{2(\bar{\theta}_1 \gamma^\mu \theta')p_\mu}}{p^2} \frac{e^{-2(\bar{\theta}'' \gamma^\mu \theta')(p-q-k)_\mu}}{|p-q-k|^{2\Delta}} \frac{e^{-2(\bar{\theta}'' \gamma^\mu \theta')q_{\mu}}}{q^{2\Delta}}
\frac{e^{-2(\bar{\theta}'' \gamma^\mu \theta')k_{\mu}}}{k^{2\Delta}}
\frac{e^{2(\bar{\theta}'' \gamma^\mu \theta_2)p_\mu}}{p^{2\Delta}}.  \label{DSEN2}
\end{align}
\\
As we did in the $\mathcal{N}=1$ case we neglect (l.h.s.) of (\ref{DSEN2}) in the IR limit and we get
\\
\begin{align} \label{DSEIRN2}
&-\frac{e^{2(\bar{\theta}_1 \gamma^\mu \theta_2)p_\mu}}{p^2}=2A_2^4 \lambda^2  \int \frac{d^d q}{(2\pi)^d} \frac{d^d k}{(2\pi)^d}
\frac{\mathcal{J}_{\mathcal{N}=2}(p,\bar{\theta}_1,\theta_2)}{p^2|p-q-k|^{2\Delta}q^{2\Delta}k^{2\Delta}p^{2\Delta}},\\
&\mathcal{J}_{\mathcal{N}=2}(p,\bar{\theta}_1,\theta_2)=\int d^2 \theta' d^2 \bar{\theta}'' e^{2(\bar{\theta_1} \gamma^\mu \theta')p_\mu} e^{-2(\bar{\theta}'' \gamma^\mu \theta')p_\mu} e^{2(\bar{\theta}'' \gamma^\mu \theta_2)p_\mu}.
\end{align}
\\
The quantity $\mathcal{J}_{\mathcal{N}=2}(p,\bar{\theta}_1,\theta_2)$, which again do not depend on $k$ and $q$, can be computed straightforwardly following the same procedure that we used for $\mathcal{J}(p,\theta_1,\theta_2)$ and $\mathcal{J}_2(p,\theta_1,\theta_2)$ (see appendix \ref{appendixB}) and one gets\footnote{Notice that in the exponent there is a different sign and $(1\leftrightarrow2)$ are inverted because we have written the DSE for $\langle \bar{\Phi}(-p,\bar{\theta}_1) \Phi(p,\theta_2) \rangle=\langle \Phi(p,\theta_2) \bar{\Phi}(-p,\bar{\theta_1}) \rangle$.}
\\
\begin{equation}
\mathcal{J}_{\mathcal{N}=2}(p,\bar{\theta}_1,\theta_2)=-4 p^2 e^{2(\bar{\theta}_1 \gamma^\mu \theta_2)p_\mu}.
\end{equation}
\\
Substituting $\mathcal{J}_{\mathcal{N}=2}(p,\bar{\theta}_1,\theta_2)$ into equation (\ref{DSEIRN2}) we get 
\\
\begin{equation}
-\frac{e^{2(\bar{\theta}_1 \gamma^\mu \theta_2)p_\mu}}{p^2}=-8A_2^4 \lambda^2 \frac{e^{2(\bar{\theta}_1 \gamma^\mu \theta_2)p_\mu}}{p^{2\Delta}}  \int \frac{d^d q}{(2\pi)^d} \frac{d^d k}{(2\pi)^d}
\frac{1}{|p-q-k|^{2\Delta}q^{2\Delta}k^{2\Delta}}.
\end{equation}
\\
The remaining integrals over momenta are equal to those in the $\mathcal{N}=1$ case. In the end we get
\begin{align}
&\Delta=\frac{d+1}{4},\qquad\qquad A_2^4=\frac{1}{8}\frac{(4 \pi)^d}{\lambda^2}\frac{\Gamma^3( \frac{d+1}{4} ) \Gamma ( 3 \frac{d-1}{4} )}{\Gamma^3 ( \frac{d-1}{4} ) \Gamma( \frac{3-d}{4} )}=\frac{1}{8}A.
\end{align}
\\
The parameter $\Delta$ is exactly equal to the one we found in the $\mathcal{N}=1$ model. Thus the scaling dimension for chiral (or anti-chiral ) fields is again $\Delta_{\Phi}=\frac{d-1}{4}$ and in $d=3-\varepsilon$ the anomalous dimension is again $\gamma_{\Phi}=\frac{\varepsilon}{4}$.

 It is worth noticing, as already pointed out in \cite{Klebanov:2016xxf}, that the superconformal algebra with $\mathcal{N}=2$ in $d=3$ fixes the dimension of chiral (or anti-chiral) operators. Chiral multiplets satisfy a shortening condition that fixes their dimension in term of the $R$-charge. Notice that the $R$-charge is fixed by the request of the invariance of the superpotential term $\int d^2 \theta W[\theta]$, which requires $R_{W[\Phi]}=2$ and $R_{\Phi}=\frac{1}{2}$. Hence holds the relation \cite{Bobev}: $\Delta_{\Phi}=\frac{d-1}{2}R_{\Phi}=\frac{d-1}{4}$.
\subsection{Bilinear singlet operators}
We now study the spectrum of singlet bilinear operators. Possible (singlet) bilinear operators are
\\
\begin{align}
	&\Phi_{abc} \DAlambert^h \Phi_{abc} & \Phi_{abc} \DAlambert^hD^2\Phi_{abc}\nonumber \\
	&\bar{\Phi}_{abc}\DAlambert^h\bar{\Phi}_{abc} & \bar{\Phi}_{abc}\DAlambert^h\bar{D}^2 \bar{\Phi}_{abc} \label{OphiBarphi} \\
	& \bar{\Phi}_{abc} \DAlambert^h \Phi_{abc} &\bar{\Phi}_{abc}\DAlambert^hD^2\Phi_{abc} \nonumber
\end{align}
\\
All others possibilities can be obtained applying $D^2$ or $\bar{D}^2$ and by complex conjugation, keeping in mind that: $D^2\bar{D}^2=\DAlambert $ , $D_\alpha D^2=0$ and $\bar{D}_{\alpha} \bar{D}^2=0$. Operators which involve only chiral or anti-chiral fields do not renormalize (there are no melonic contributions in the Large $N$ limit) and the only remaining operators are those in the last line of (\ref{OphiBarphi}).
Since $D_\alpha\bar{\Phi}=0$, we can write $\bar{\Phi}D^2\Phi$ as a (super-)descendants of $\bar{\Phi}\Phi$: $D^2\left( \bar{\Phi} \Phi \right)=\bar{\Phi}D^2\Phi$. Thus it is sufficient to study the three point function of $\bar{\Phi}\DAlambert^h{\Phi}$ and find its scaling dimension, then also the scaling dimensions of its descendants are fixed. 

\begin{figure}[tb!]
	\centering
	\includegraphics[width=0.4\linewidth]{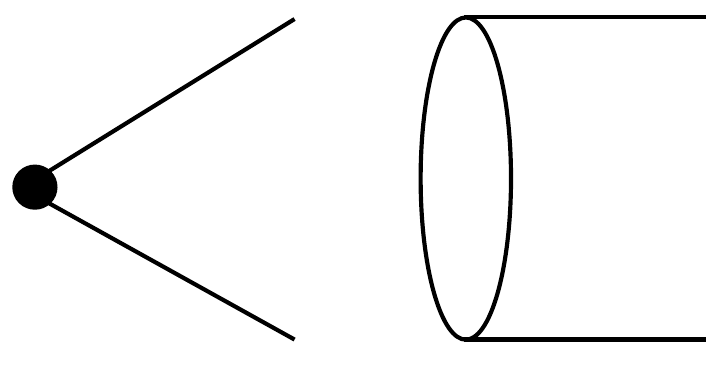}
	\caption{Radiative corrections to the three point function.}
	\label{fig:appliedkernel}
\end{figure}

Thinking in terms of diagrams, melonic dominance constrains the radiative corrections of 3pt function to be of the form in Figure~\ref{fig:appliedkernel}.
\\
\\
Contributions of this form can be non vanishing only if the Kernel
operator consists of a chiral vertex and an anti-chiral one (otherwise two-point-functions $\langle \Phi \Phi \rangle$ or $\langle \bar{\Phi} \bar{\Phi} \rangle$ in the loop vanish).

Our ansatz for the three point function of ${\mathcal O}\equiv\bar{\Phi}_{abc} \Phi_{abc}$ with a chiral and an anti-chiral field is
\\
\begin{equation}
V(p,\theta_1,\bar{\theta}_2)=\langle {\mathcal O}(0) \Phi(-p,\theta_1)\bar{\Phi}(p,\bar{\theta_2})\rangle=\frac{e^{-2(\bar{\theta}_2 \gamma^\mu \theta_1)p_\mu}}{p^{2\Delta+\Delta_{\mathcal O}}}
\end{equation} 
\\
The eigenvalue equation takes the form
\begin{align} \nonumber 
&\left(\mathcal{K}_{\mathcal{N}=2}V\right)(p,\theta_1,\bar{\theta}_2)+6 \lambda^2 A_{\mathcal{N}=2}^4  \int d^2 \bar{\theta}' d^2 \theta'' \int \frac{d^d q}{(2\pi)^d} \frac{d^d k}{(2\pi)^d}
\\
&\frac{e^{-2(\bar{\theta}'\gamma^\mu \theta_1)p_\mu}}{p^{2\Delta}} \frac{e^{-2(\bar{\theta}_2\gamma^\mu \theta'')p_\mu}}{p^{2\Delta}} \frac{e^{2(\bar{\theta}'\gamma^\mu \theta'')p_\mu}}{q^{2\Delta}|p-q-k|^{2\Delta+\Delta_{\mathcal O}}k^{2\Delta}}.
\end{align}
\\
Noticing that the contribution of the integration over Grassmann variables is precisely equal to $\mathcal{J}_{\mathcal{N}=2}(p,\bar{\theta}_1,\theta_2)$, we get
\\
\begin{equation} \label{N2KernelMom}
\left(\mathcal{K}_{\mathcal{N}=2}V\right)(p, \theta_1, \theta_2)=-8\times 3 A_{\mathcal{N}=2}^4\lambda^2 \frac{e^{-2(\bar{\theta}_2 \gamma^\mu \theta_1)p_\mu}}{p^{4\Delta-2}} \int \frac{d^d q}{(2\pi)^d}\frac{d^d k}{(2\pi)^d}\frac{1}{q^{2\Delta}k^{2\Delta+\Delta_{\mathcal O}}|p-q-k|^{2\Delta}}.
\end{equation}
\\ 
Since $8A^4_{\mathcal{N}=2}=A^4$, equation (\ref{N2KernelMom}) reduces to (\ref{N1KernelMom}) and has exactly the same solutions
\\
\begin{equation} \label{FinalResult}
g_{\mathcal{N}=2}(\Delta_{\mathcal O})=-3\frac{\Gamma(\frac{d+1}{4})\Gamma(3\frac{d-1}{4})\Gamma(\frac{d-1}{4}-\frac{\Delta_{\mathcal O}}{2})\Gamma(\frac{3-d}{4}-\frac{\Delta_{\mathcal O}}{2})}{\Gamma(\frac{d-1}{4})\Gamma(\frac{3-d}{4})\Gamma(\frac{d+1}{4}+\frac{\Delta_{\mathcal O}}{2}){\Gamma(3\frac{d-1}{4}-\frac{\Delta_{\mathcal O}}{2})}}=1.
\end{equation}
\\
The lowest scaling dimension in  $d=3-\varepsilon$ is
\\
\begin{equation}
	\Delta_{\bar{\Phi}\Phi}=1+\varepsilon+3\varepsilon^2+O(\varepsilon^3)
\end{equation}
\\
and its (super-)descendant $\bar{\Phi}D^2\Phi$ has dimension
\\
\begin{equation}
\Delta_{\bar{\Phi}D^2\Phi}=\Delta_{\bar{\Phi}\Phi}+1=2+\varepsilon+3\varepsilon^2+O(\varepsilon^3).
\end{equation}

The expression \eqref{FinalResult} is well defined for all values $1<d<3$, while is singular at the two extremes.

\subsection{Spinning bilinear operators}
\label{sec:spinning}
We now study the spectrum of spinning (with integer spin) bilinear operators of the form
\begin{equation} \label{GuessSpinning}
	\mathcal O^{(\ell)}=\bar{\Phi}_{abc} \partial_{\mu_1}..\partial_{\mu_\ell} \DAlambert^h \Phi_{abc}.
\end{equation}
Working in the momentum space, our ansatz for the three point function is
\\
\begin{equation} \label{spinning3ptfunction}
	V^\ell_{\mu_1..\mu_\ell}(p,\theta_1,\bar{\theta}_2)=\langle \mathcal O^{(\ell)}(0) \Phi(-p,\theta_1)\bar{\Phi}(p,\bar{\theta_2})\rangle=\frac{e^{-2(\bar{\theta}_2 \gamma^\mu \theta_1)p_\mu}}{p^{2\Delta+\Delta_\ell+\ell}}p_{\mu_1}...p_{\mu_\ell}.
\end{equation} 
Diagrams contributing to the three-point functions have exactly the same structure as those contributing to the scalar bilinear. The only difference is how     $	V^\ell_{\mu_1..\mu_\ell}(p,\theta_1,\bar{\theta}_2)$
depends on momenta: 
\begin{equation} \label{spinningKernel}
	\left(\mathcal{K}_{\mathcal{N}=2}V^\ell_{\mu_1..\mu_\ell}\right)(p, \theta_1, \theta_2)=-3 A^4\lambda^2 \frac{e^{-2(\bar{\theta}_2 \gamma^\mu \theta_1)p_\mu}}{p^{4\Delta-2}} \int \frac{d^d q}{(2\pi)^d}\frac{d^d k}{(2\pi)^d}\frac{k_{\mu_1}...k_{\mu_\ell}}{q^{2\Delta}k^{2\Delta+\Delta_\ell}|p-q-k|^{2\Delta}}.
\end{equation}
\\
In order to solve equation (\ref{spinningKernel}) we can contract each $k_\mu$ with an arbitrary null-vector $\xi^\mu$
\\
\begin{align}\nonumber
	\left(\mathcal{K}_{\mathcal{N}=2}V^\ell_{\mu_1..\mu_\ell}\right)(p, \theta_1, \theta_2) \xi^{\mu_1}..\xi^{\mu_\ell}=&-3 A^4\lambda^2 \frac{e^{-2(\bar{\theta}_2 \gamma^\mu \theta_1)p_\mu}}{p^{4\Delta-2}} \\
	&\int \frac{d^d q}{(2\pi)^d}\frac{d^d k}{(2\pi)^d}\frac{(k \cdot \xi)^\ell}{q^{2\Delta}k^{2\Delta+\Delta_\ell}|p-q-k|^{2\Delta}}.
\end{align}
\\
Now the integration can be performed by means of the known integral \cite{GiombiKleb}
\\
\begin{align}
	&\int d^d k \frac{\left( z \cdot k \right)^\ell}{k^{2\alpha}(k-p)^{2\beta}}=L_{d,\ell}(\alpha,\beta)\frac{\left( z \cdot p \right)^\ell}{(p^2)^{\alpha + \beta-\frac{d}{2}}},\\
	& L_{d,\ell}(\alpha,\beta)=\pi^{\frac{d}{2}} \frac{\Gamma(\frac{d}{2}-\alpha+\ell) \Gamma(\frac{d}{2}-\beta) \Gamma(\alpha+\beta-\frac{d}{2})}{\Gamma(\alpha) \Gamma(\beta) \Gamma(d+\ell-\alpha-\beta)}.
\end{align}
\\
The final result for the eigenvalue $g_{\mathcal{N}=2}(\Delta_\ell,\ell)$ is
\\
\begin{equation} \label{spinningEigenvalue}
	g_{\mathcal{N}=2}(\Delta_\ell,\ell)=-(-1)^\ell 3\frac{ \Gamma \left(\frac{3 (d-1)}{4}\right) \Gamma \left(\frac{d+1}{4}\right) \Gamma \left(\frac{d-1}{4}+\frac{\ell-\Delta_\ell}{2}-\right) \Gamma \left(\frac{3-d}{4}+\frac{\Delta_\ell+\ell}{2}\right)}{\Gamma \left(\frac{3-d}{4}\right) \Gamma \left(\frac{d-1}{4}\right) \Gamma \left(\frac{3 (d-1)}{4}+\frac{\ell-\Delta_\ell}{2}\right) \Gamma \left(\frac{d+1}{4}+\frac{\Delta_\ell+\ell}{2}\right)}=1.
\end{equation}
\\

\begin{figure}[t]
	\centering
	\includegraphics[width=0.6\linewidth]{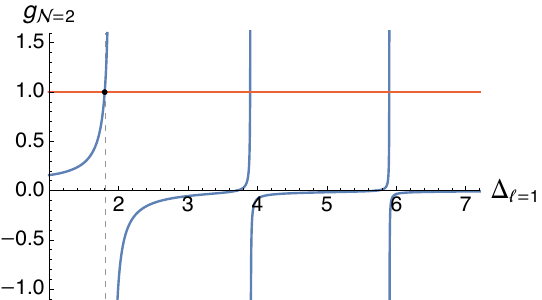}
	\caption{Solutions of eigenvalue problem for scalar bilinears of spin-1 in $d=2.8$ dimensions. The vertical dashed line corresponds to the unitarity bound. The black point represents the stress tensor multiplet.}
	\label{plotgbb}
\end{figure}
It can be checked that setting $\ell=0$, equation~\eqref{spinningEigenvalue} reduces to results for scalar operators (\ref{FinalResult}).
A nice consistency check is that in any $d$ there exists a solution corresponding to the conserved stress-energy tensor, as shown in Figure~\ref{plotgbb}.
The stress-energy tensor sits in a super-conformal multiplet with the bottom component being a spin $1$ operator \cite{CordovaInt}. Therefore we expect a solution for $\ell=1$, $\Delta_\ell=d-1$ and any $d$.
Substituting the mentioned quantum numbers into (\ref{spinningEigenvalue}) it is easy to check that the expected solution does exists
\begin{equation} \label{stresEnergy}
	g_{\mathcal{N}=2}(d-1,1)=1.
\end{equation}
Setting $h=0$ in (\ref{GuessSpinning}), we can parametrize the solution of (\ref{spinningEigenvalue}) as: $\Delta_{\ell}=2\Delta_{\Phi}+\ell+\gamma_\ell$. In $3-\varepsilon$ dimensions we find 
\begin{align}
	&\Delta_\ell=2\Delta_{\Phi}+\ell+\frac{3 (-1)^\ell  }{4 \ell+2}\varepsilon + \\
	&\frac{3 (-1)^\ell \left((2 \ell+1) \left((2 \ell+1) H_{\ell-\frac{1}{2}}+2 \ell (\log (4)-3)-1+\log (4)\right)-3 (-1)^\ell\right)}{4 (2 \ell+1)^3} \varepsilon ^2 \nonumber
\end{align}
where $H_{\ell-\frac{1}{2}}$ are Harmonic numbers. Substituting $\ell=1$ we find $\Delta_1=2-\varepsilon$ which correspond to the stress-energy tensor consistently with (\ref{stresEnergy}).

\begin{figure}[t]
	\centering
	\includegraphics[width=0.7\linewidth]{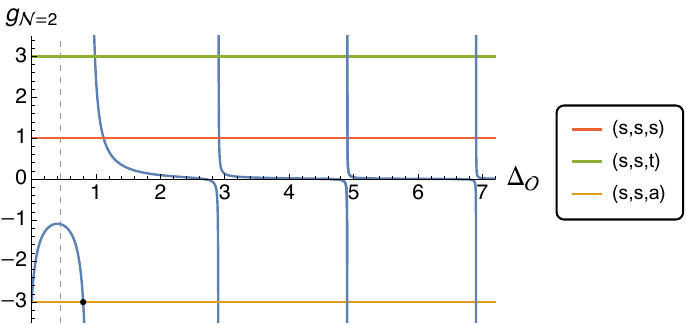}
	\caption{Solutions of eigenvalue problem for scalar bilinears in various representations in $d=2.8$ dimensions. The black dots corresponds the the global symmetry current multiplet. The vertical dashed line corresponds to the unitarity bound.}
	\label{plotAllIrreps}
\end{figure}

\subsection{Non singlet bilinears}
\label{sec:nonsinglet}

In this section we extend the computation of operator dimension to non-singlet bilinears of the form $\mathcal O  = \bar \Phi\Box^h \Phi$. Depending on whether the indexes of the $O(N)^3$ symmetry are contracted, symmetriezed or anti-symmetrized, we will have various representations $(r_1,r_2,r_3)$:
\be
\label{eq:OPEON3}
(v,v,v)\otimes (v,v,v)\sim & \,\,(s,s,s) \oplus (s,s,t)  \oplus (s,s,a)  \oplus (s,t,t)  \oplus (s,t,a) \nonumber\\
& \oplus (s,a,a)  \oplus (t,t,t)  \oplus (t,t,a)  \oplus (t,a,a)  \oplus (a,a,a) \,, 
\ee
Also, given the permutation symmetry of the indexes, the actual order of $r_i$ will not matter. In order to compute the DSE satisfied by the three point function $\Phi \bar{\Phi} \mathcal O_{(r_1,r_2,r_3)}$, we can simply compute the insertion of a ladder in the three point function as in Figure~\ref{fig:appliedkernel}.

\begin{table}[ht]
\centering
\begin{tabular}{@{}c|c|c@{}}
	\toprule
$(r_1,r_2,r_3)$ & three-level & $+1$ ladder \\
	\midrule
	\midrule
$(s,s,s) $& $\frac1{\sqrt{N^3}} $& $\frac32 g^2 \left( 2+2N +N^3\right)$\\
\midrule
$(s,s,t) $&$ \frac1N $& $\frac12 g^2 \left( 6+7N +N^3\right)$\\
\midrule
$(s,s,a) $&$ \frac1N $& $-\frac12 g^2 \left( 2+3N +N^3\right)$\\
\midrule
$(s,t,t) $&$ \frac1{\sqrt N}$ & $\frac12 g^2 \left( 6+5N \right)$\\
\midrule
$(s,t,a) $&$ \frac1{\sqrt N } $& $-\frac12 g^2 \left( 2+N \right)$\\
\midrule
$(s,a,a)$ & $\frac1{\sqrt N } $& $\frac12 g^2 \left( 2+N \right)$\\
\midrule
$(t,t,t)$ & $ 1 $& $\frac32 g^2 \left( 2+N \right)$\\
\midrule
$(t,t,a)$ & $ 1 $& $\frac12 g^2 \left(- 2+N \right)$\\
\midrule
$(t,a,a)$ & $ 1 $& $-\frac12 g^2 \left( 2+N \right)$\\
\midrule
$(a,a,a)$ & $ 1 $& $-\frac32g^2 \left( -2+N \right)$\\
\bottomrule 
\end{tabular}
	\caption{Coefficient multiplying the group tensor structure and the spacetime factor for each three point function $\langle \Phi \bar{\Phi} \mathcal O_{(r_1,r_2,r_3)}\rangle$. The second row is the overall three level coefficient. The third row must be intended as a correction and should be multiplied by the three level coefficient to get the correct factor. The bilinear normalization is chosen such that their 2pt function is $N$-independent.}
	\label{tab:results}
\end{table}

In Table~\ref{tab:results} we report the results for each representation. We see that in the large-$N$ limit, with $g^2 N^3$ kept fixed, only the first three representations have a non trivial DSE. Moreover, the only difference between the three representation is a $\pm 3$ factor. Hence, given the DSE~\eqref{FinalResult}, we have:

\begin{eqnarray}
\label{eq:DSEnonsinglet}
&(s,s,s): \qquad & g_{\mathcal{N}=2}(\Delta_{\mathcal O})= 1, \nonumber \\
&(s,s,a): \qquad & g_{\mathcal{N}=2}(\Delta_{\mathcal O})= -3, \nonumber \\
&(s,s,t): \qquad & g_{\mathcal{N}=2}(\Delta_{\mathcal O})= 3. \nonumber \\
\end{eqnarray}

As sanity check, one can verify that the eigenvalue equation for the representation $(s,s,a)$ contains the solution $\Delta_{\mathcal O} = d-2$, which correspond to the scalar multiplet associated to the global symmetry currents of $O(N)^3$. We recall indeed that in $\mathcal N=2$ superconformal field theries, the conserved global symmetry currents sits in a scalar real superfield with dimension exactly equal to $d-2$. We show the eigenvalue equations~\eqref{eq:DSEnonsinglet} in Figure~\ref{plotAllIrreps}.

At this point, it is straightforward to combine the results of section~\ref{sec:spinning} and this section to get the spectrum of spinning bilinears in any representation. In particular one can check that there are no other conserved multiplets in non-singlet representation, as there are no other scalars of dimension $d-2$ in non-adjoint representations.

\subsection{Perturbative checks}
In this section we study the model perturbatively in a $3-\varepsilon $ expansion and we start our analysis by revising the results in \cite{POP1}
\begin{align}  \nonumber
	&S[\Phi,\bar{\Phi}]= \int d^dx d^2\theta d^2\bar{\theta} ( \Phi_{abc} \bar{\Phi}_{abc}) +\int d^d y d^2 \theta \left( \frac{1}{4} g_t {\mathcal O}_t + \frac{1}{4} g_p {\mathcal O}_p+\frac{1}{4} g_{dt} {\mathcal O}_{dt} \right)+\\ 
	& \int d^d \bar{y} d^2 \bar{\theta} \left( \frac{1}{4} g_t \bar{{\mathcal O}}_t + \frac{1}{4} g_p \bar{{\mathcal O}}_p+\frac{1}{4} g_{dt} \bar{{\mathcal O}}_{dt} \right) \\ 
&	\begin{cases}
{\mathcal O}_t=\Phi_{a_1 b_1 c_1} \Phi_{a_1b_2c_2} \Phi_{a_2b_1c_2}  \Phi_{a_2b_2c_1}\\ 
 {\mathcal O}_p=\frac{1}{3}\left(\Phi_{a_1 b_1 c_1} \Phi_{a_2 b_1 c_1} \Phi_{a_1 b_2 c_2} \Phi_{a_2 b_2c_2}+\text{perm} \right)\\ 
{\mathcal O}_{dt}=\Phi_{a_1 b_1 c_1} \Phi_{a_1b_1c_1} \Phi_{a_2b_2c_2} \Phi_{a_2b_2c_2}
	\end{cases}
\end{align}
\\
in which we included all the possible $O(N)^3$ invariant interactions. The reason is that when doing perturbation theory one should expect that radiative corrections generate all the possible invariants. 

The beta-functions can be found in \cite{Gracey} and receive corrections only by field renormalization 
\\
\begin{equation}
\beta_t=\left(-\varepsilon+4\gamma_{\Phi}\right) g_t,  \ \ \ \ \ \
 \beta_p=\left(-\varepsilon+4\gamma_{\Phi}\right) g_p,  \ \ \ \ \ \
\beta_{dt}=\left(-\varepsilon+4\gamma_{\Phi}\right) g_{dt}
\end{equation}
\begin{align}
&\gamma_{\Phi}=\frac{1}{6\pi}( 12g_t g_p(1+N+N^2)+6g_{dt}^2(2+N^3)+3g_t^2(2+3N+N^3)+\nonumber\\
&\phantom{----}g_p^2(5+9N+3N^2+N^3) + 36 g_t g_{dt}N+12g_p g_{dt}(2+N+N^2) )
\end{align}
\\
Defining scaled parameters $\lambda_t$, $\lambda_p$ and $\lambda_{dt}$ as 
\begin{equation}\label{ScaledPar}
	g_t=\frac{\pi}{\sqrt{2}}\frac{\lambda_t}{N^{\frac{3}{2}}}, \ \ \ \ \ \ g_p=\frac{\pi}{\sqrt{2}}\frac{\lambda_p}{N^{\frac{5}{2}}}, \ \ \ \ \ \ g_{dt}=\frac{\pi}{\sqrt{2}}\frac{\lambda_{dt}}{N^{\frac{7}{2}}},
\end{equation}
we then obtain the anomalous dimension
\begin{equation}
	\gamma_\Phi(\lambda_t,\lambda_p,\lambda_{dt})=\frac{\lambda_t^2}{4}\,,
\end{equation}
\\
and scaled beta-functions 
\\
\begin{equation} \label{ScaledParameters}
\beta_{\lambda_t}=\left(-\varepsilon+4\gamma_{\Phi}\right) \lambda_t, \ \ \ \ \ \ \ \beta_{\lambda_p}=\left(-\varepsilon+4\gamma_{\Phi}\right) \lambda_p, \ \ \ \ \ \ \ 
\beta_{\lambda_{dt}}=\left(-\varepsilon+4\gamma_{\Phi}\right) \lambda_{dt}\,.
\end{equation}
\\
It is interesting that beta-function (\ref{ScaledParameters}) do not fix $\lambda_p$ and $\lambda_{dt}$ at the fixed point. In practice there exist a continuum of fixed points for $\lambda_t^2=\varepsilon$ and arbitrary $\lambda_p$ and $\lambda_{dt}$.
The anomalous dimension at the fixed point matches with the result of DSE: $\gamma_{\Phi}=\frac{\varepsilon}{4}$ (in $3-\varepsilon$ dimensions).\\
We can now study matrix of derivatives of beta-functions $\left(\frac{\partial \beta}{\partial \lambda}\right)_{ij}=2\lambda_t \lambda_j \delta_{t j}+ (-\varepsilon+\lambda_t^2)\delta_{ij}$. Evaluated at the fixed point we get
\\
\begin{gather}
\left(\frac{\partial \beta}{\partial \lambda}\right)_{ij}=\begin{pmatrix} 2\varepsilon & 0 & 0\\ 2\varepsilon^{\frac{1}{2}}\lambda_p & 0 & 0\\ 2\varepsilon^{{\frac{1}{2}}} \lambda_{dt} & 0 & 0
\end{pmatrix}.
\end{gather}
\\
The theory turns out to be marginally stable because of the presence of two marginal operators in the super-potential, corresponding to the two null eigenvalues of the matrix.

The very non trivial part of the results is that we have found one non zero eigenvalue. Chiral and anti-chiral super potential should be stable quantities and we did not expect any anomalous dimension at the fixed point.
One possibility to interpret the results is that at the fixed point, a multiplet recombination must happen. Indeed in the UV there is a $U(1)$ global symmetry that rotates $\Phi$ and $\bar{\Phi}$. The conservation of the current associated to this symmetry can be expressed by the constraint $D^2\left( \bar{\Phi} \Phi\right)=0$. In the IR the super-potential $W[\Phi]$ breaks the symmetry and the current is no more conserved. We expect at the fixed point $D^2 \left( \bar{\Phi}\Phi \right)=B$, being $B$ an operator that breaks the $U(1)$ symmetry. The tetrahedral interaction has the right dimension and R-charge so we guess
\\
\begin{equation}
\Phi_{a_1 b_1 c_1} \Phi_{a_1b_2c_2} \Phi_{a_2b_2c_2} \Phi_{a_2b_2c_1} \propto D^2 \left( \bar{\Phi} \Phi \right)
\end{equation}
\\
meaning that it becomes a super-descendant of the $\bar{\Phi} \Phi$ operator. Supporting our assumption we find the scaling dimension of $\mathcal O_{t}$
\begin{equation}
	\Delta_{tetra}=2-\varepsilon+2\varepsilon+O(\varepsilon^2)=2+\varepsilon+O(\varepsilon^2)
\end{equation}
\\
that matches with the dimension of $\bar{\Phi}D^2 \Phi$ found solving the kernel equation.

\subsection{Large spin expansion}

In section \ref{sec:nonsinglet} we showed that certain bilinear operators do not have the naive dimension  $2\Delta_\Phi$, but  acquire an anomalous dimension, even at infinte $N$. Given the well known results about  dimensions of bilinear operators  (i.e. double trace) in a CFT \cite{Komargodski:2012ek,Fitzpatrick:2012yx,Alday:2016njk,Alday:2016jfr,Caron-Huot:2017vep}, it is instructive to check that the solution we found follows the expected behaviour at large spin $\ell$. 

\begin{figure}[t!]
		\subfigure[]{\includegraphics[width=0.48\textwidth]{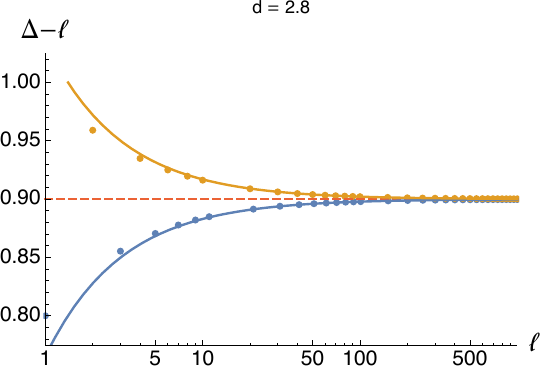}\label{fig:Regge2.8}}  \quad \ \
		\subfigure[]{\includegraphics[width=0.48\textwidth]{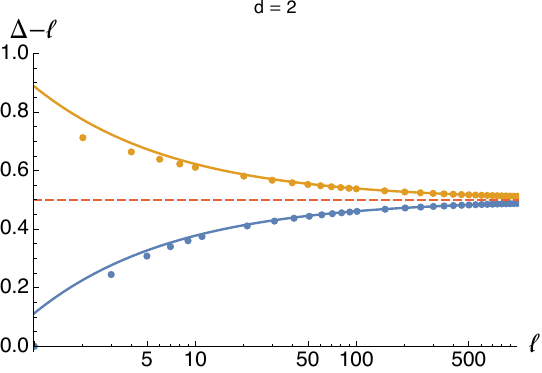}\label{fig:Regge2.0}}
		\caption{Twist of the first two families of singlet bilinear operators as a function of the spin $\ell$ for two different values of spacetime $d$. The two families correspond to spin odd (blue) and spin even (yellow). The dots represents numerical solutions of the eigenvalue equation \eqref{spinningEigenvalue}. We did not report all spins. The continuous lines show the leading correction given in  \eqref{eq:largespin}. The dots for $\ell=1$ correspond to the supermultiplet of the stress tensor.} 
		\label{fig:Regge}
\end{figure}

Hence, we start from the eigenvalue equation \eqref{spinningEigenvalue} and substitute the ansatz $\Delta_{\mathcal O} = 2\Delta_\Phi + \ell +\gamma$. Here we focus on the leading double trace trajectory with $h=0$. Assuming that the correction $\gamma$ is suppressed at large $\ell$, we first expand for small $\gamma$. This produces a simple pole, with a residue that should be equal to the eigenvalue $e=1,\pm 3$, depending on the representation considered. Then, expanding for large $\ell$ we obtain
\be
\label{eq:largespin}
\gamma \simeq \frac1e \frac{6 (-1)^\ell  \Gamma\left(\frac{3 (d-1)}{4}\right) \Gamma\left(\frac{d+1}{4}\right)}{\Gamma
 \left(\frac{3-d}{4}\right) \Gamma \left(\frac{d-1}{4}\right) \Gamma\left(\frac{d-1}{2}\right)} \frac{1}{\ell^{2\Delta_\Phi}} + \ldots
\ee
In Figure~\ref{fig:Regge} we show the twist $\tau=\Delta-\ell$ of the leading solutions of the singlet DSE as a function of $\ell$. Notice that even and odd spins organize in distinct families, both approaching $\tau=2\Delta_\Phi$ at large $\ell$. Also, the  spin-$1$ belongs to the lower family. We also show the correction found in \eqref{eq:largespin}, which nicely fits the dimensions even at intermediate values of $\ell$ 
 
Alternatively one can explore the large spin behavior in $d=3-\varepsilon$ dimensions. Plugging $\Delta_{\mathcal O}= 2\Delta_\Phi + \ell +\varepsilon \tilde \gamma$ and expanding at leading oder in $\varepsilon$ we obtain
\be
\label{eq:epsilonlargeL}
\Delta_{\mathcal O}= 2\Delta_\Phi + \ell + \frac32 \frac{(-1)^\ell}{2\ell+1} \varepsilon \xrightarrow{\,\,\ell=1\,\,} 2-\varepsilon\,.
\ee

Notice that the above expression is exact at $O(\varepsilon)$: it resums the whole large spin perturbative series and connects the stress tensor multiplet at $\ell=1$ to the large spin value of the twist $2\Delta_\Phi$. Also in this equation~\eqref{eq:epsilonlargeL}, the leading order correction is $\ell^{-2\Delta_\Phi}$.

A few comments about the result in \eqref{eq:largespin} are in order. First we note that in $O(N)^3$ tensor models certain families of bilinear operators acquire an anomalous dimension already at infinite $N$, contrarily to what happens in other large $N$ theories such as $O(N)$ vector models. In the latter theories all bilinears have the naive dimension $2\Delta_\phi$.  

A second interesting fact is that, according to the lightcone bootstrap approach, the correction shown in equation~\eqref{eq:largespin} should be determined by the operator with the lower twist exchanged in the crossed channel of the 4pt function of $\Phi$. On the other hand, those families of operators, when  $N$ is strictly infinite, are decoupled from the theory, i.e. they do not appear in any correlation function. For the 4pt function $\langle\Phi\bar\Phi\Phi\bar\Phi\rangle$, this is evident  from the scaling of 3pt point functions reported in Table~\ref{tab:results}. Indeed, when we solve for the eigenfunctions of the 3pt functions DSE, we only get the allowed  dimensions of bilinear operators, but we don't get any information concerning the OPE coefficient, which might be zero. It seems paradoxical then that the double trace operators acquire an anomalous dimension despite being completely decoupled from the theory.

The resolution of the tension is to realize that the CFT here described is not a standalone theory, but it is obtained as a large-$N$ limit, and should therefore admit a perturbative expansion in $1/N$.  Imposing the consistency of the perturbative expansion order by oder, one can probe the theory in a regime where bilinear operators couple with the rest of the theory. A careful inspection of the crossing equations for $O(N)^3$ in the lightcone limit shows that the $\ell^{-2\Delta_\Phi}$ correction is indeed expected and consistent \cite{toappear}. 

\newpage
\section{Conclusions}

The melonic dominance observed in SYK models and tensor models in the large-$N$ limit represents a promising way to study quantum field theories in a strongly interacting regime, and yet be able to perform exact computations. In this limit, it is possible to write simple Dyson-Schwinger equations for 2 and 3pt functions and solve them by doing a conformal ansatz. 

We reviewed the results of \cite{POP1} for a tensor model with minimal supersymmetry and applied similar techniques to the case of four supercharges. Since the structure of the supersymmetry algebra in this case  is similar to the case of $\mathcal{N}=1$ supersymmetry in $d=4$, we could use fundamental results such as analytic structure of superspace and non-renormalization theorems.
This is particularly clear in the structure of scalar bilinear operators:  although it is possible to construct several bilinear operators, in presence of four supercharges we need to study only the operator $\bar{\Phi} \Phi$. All other types either are super-descendants or do not receive radiative corrections in the melonic limit.\\
In addition, we verified the occurrence of multiplet recombinations. In particular, the scalar singlet $\bar\Phi\Phi$ receives quantum corrections and is lifted from the supersymmetric unitarity bound.\footnote{In the free theory this supermultiplet would be associated to an extra $U(1)$ global current that is broken by the interactions.} Therefore, it must recombine with a chiral multiplet,  according to the recombination rule  \cite{CordovaInt} \mbox{$A_2 \bar{A}_2 \oplus L\bar{B}_1 \oplus B_1 \bar{L} \rightarrow L\bar{L}$}.
The eaten chiral field can be identified with a combination of the three quartic scalar singlets obtained contracting four fields $\Phi_{abc}$ in different ways. We explicitly checked this mechanism in $3-\varepsilon$ dimension: out of the three scalar singlets, two combinations remain superprimaries and their dimension is exactly $4\Delta_\Phi$, while a third gets the right anomalous dimension to become a superdescendant.

We also computed the spectrum of  bilinears with spin and bilinears transforming in all irreps of $O(N)^3$ appearing in the OPE of two fundamental representations. This allowed us to check the presence in the spectrum of the $R$-symmetry supercurrent in the spin-1 singlet sector and the $O(N)^3$ global current multiplet in the spin-0 adjoint sector.  All other sectors do not contain conserved operators. We also showed that only the bilinears in the irrep $(s,s,s), (s,s,a)$ and $(s,s,t)$ have dimension different from $2\Delta_\Phi$.

In addition, we initiated a study of the large spin behavior of double trace operators in tensor models. These models represent a great opportunity to test analytic bootstrap techniques and large spin perturbation theory in CFTs. \\
Contrarily to more familiar vector models, the dimension of the fundamental field does not approach the unitarity bound $d/2-1$. Instead $\Delta_\Phi=(d-1)/4$ in the present model. Hence, the twist $\tau=\Delta-\ell$ of the leading Regge trajectory must necessarily increase from the minimal value of, say, the stress tensor $\tau_{s.t.}= d-2$ to the asymptotic value $\tau_{asym}= (d-1)/2$.  We showed two such trajectories in Figure~\ref{fig:Regge}. Similar behaviour is found in other trajectories as well.  In vector models instead one finds a constant twist for all double traces.  \\ \indent
Moreover, we computed the leading correction at large spin for generic $d$ and compared with the exact solution, finding good agreement down to $\ell\sim10$. It would be interesting to perform a systematic analysis of tensor models  along the lines of \cite{Alday:2019clp,Henriksson:2020fqi}. We leave this direction for future investigations.

Finally, it is worth noticing that all our result cannot be extended to $d=3$, because the eigenvaue equation~\eqref{FinalResult} is singular in this limit. An interesting possibility to get a theory without divergences in $3d$ is to introduce a supersymmetric theory with $\Phi^3$ interaction and disorder. The disorder is needed in order to get melonic dominance and leads to similar DSE in the large $N$ limit. The main advantage of having a theory in integer dimensions is the absence of trivial violations of unitarity \cite{Hogervorst:2015akt}, which open the possibility to bootstrap the model, even at finite $N$. However one has to be careful with the loss of unitarity due to the disorder, see for instance \cite{Cardy:2013rqg, Hogervorst:2016itc}. The $\mathcal{N}=1$ case has been studied in \cite{POP1} and could be generalized to the present model. \\
On the other hand the model discussed in this work has not shown any pathology in $d=2$ giving us the opportunity to further cross-check our results. 
In fact supersymmetric tensor models in $d=2$ with 4 supercharges have already been studied in \cite{Chang:2019yug} and it is easy to verify that the numerical solutions of our (\ref{spinningEigenvalue}) when d=2 are in perfect agreement with their results.

\section*{Acknowledgments}
We thanks Johan Henriksson for illuminating discussions about large spin perturbation theory and the interpretation of our results. We also thank Marten Reehorst for collaborating on the initial stages of this project and Igor Klebanov and Andrea Manenti for useful comments. 
This project has received funding from the European Research Council (ERC) under the European Union's Horizon 2020 research and innovation programme (grant agreement no. 758903).


\appendix
\section{3d superspace} \label{appendixA}
The Lorentz Algebra in three dimensions is isomorphic to $sl(2, R)$ and its fundamental representation is real, meaning that is similar to its complex conjugate representation. The fundamental representation acts on a (real) two-component Majorana spinor $\psi^{\alpha}=(\psi^{1}, \psi^{2})$ and indices are raised and lowered by the following matrix \cite{POP1,Gracey,GRIS}

\begin{gather} \label{SpinorMetrics}
	C_{\alpha \beta}=-C^{\alpha \beta}=\begin{pmatrix} 0 & -i \\ i & 0 
	\end{pmatrix},
\end{gather}
\\
By means of (\ref{SpinorMetrics}) is also possible to define a scalar product between spinors

\begin{equation}
	\psi^{\alpha}\chi_{\alpha}\equiv \psi^{\alpha} C_{\alpha \beta} \chi^{\beta} \ \ \ \ \ \ \ \psi^2\equiv\frac{1}{2}\psi^{\alpha}\psi_\alpha=i\psi^2\psi^1
\end{equation}
\\
The fermionic part of the supersymmetry algebra is defined as usual by the anti-commutation relation \cite{PARK1,NIZAMI}

\begin{equation} \label{NSusyAlgebra}
	\{ Q^I_\alpha,Q^J_\beta \} \ = \ 2  \gamma^\mu_{\alpha \beta}P_\mu,
\end{equation}
\\
where $I$ and $J$ range from $1$ to $\mathcal{N}$ and $Q^I_\alpha$ are Majorana spinors. Matrices $\gamma^\mu$ are the Dirac Matrices and satisfy Clifford's algebra
\\
\begin{equation} \label{CliffordAlgebra}
	\{\gamma^\mu , \gamma^\nu\}=2\eta^{\mu \nu}, \ \ \ \ \ \ \
	g^{\mu\nu}=\left(-1,1,1\right).
\end{equation}
\\
In 3d the Dirac Matrices are $2\times2$ matrices and can be chosen to be real \cite{GRIS}
\\
\begin{gather} \label{GammaMatrices}
	\gamma^0=\begin{pmatrix} 0 & -1 \\ 1 & 0 
	\end{pmatrix},
	\ \
	\gamma^1=\begin{pmatrix} 0 & 1 \\ 1 & 0 
	\end{pmatrix},
	\ \
	\gamma^2=\begin{pmatrix} 1 & 0 \\ 0 & -1 
	\end{pmatrix}.
\end{gather}
\\
It is worth noticing that gamma matrices in equation (\ref{CliffordAlgebra}) have one high index and one low index $\gamma^{(\mu) \alpha}_{\ \ \ \ \beta}$, so that contractions are trivial. When gamma matrices appear with two low indices (or high) we implicitly mean that one index has been lowered using $C_{\alpha \beta}$. We define
\\
\begin{align}
	&\gamma^{(\mu)}_{\alpha \beta}=C_{\alpha \rho} \gamma^{(\mu) \rho}_{\ \ \ \ \beta},\\
	&\gamma^{(\mu)\alpha}_{\ \ \ \ \beta}=C^{\alpha \rho} \gamma^{(\mu)}_{\rho \beta}.
\end{align}
\\
The resulting matrices can be easily computed 
\\
\begin{gather} \label{GammaMatricesLowIndices}
	\gamma^{(0)}_{\alpha \beta}=\begin{pmatrix} -i & 0 \\ 0 & -i 
	\end{pmatrix},
	\ \ \
	\gamma^{(1)}_{\alpha \beta}=\begin{pmatrix} -i & 0 \\ 0 & i 
	\end{pmatrix},
	\ \ \
	\gamma^{(2)}_{\alpha \beta}=\begin{pmatrix} 0 & i \\ i & 0 
	\end{pmatrix}
\end{gather}
\\
and they are imaginary and symmetric.  

\subsection{$\mathcal{N}=1$}
If $\mathcal{N}=1$ there is only one spinor generator $Q_\alpha$ and $2$ supercharges. Thus, in the corresponding super-space there are only two Grassmann numbers $\theta^1$ and $\theta^2$ assembled into the Majorana spinor $\theta^\alpha$. In the rest of the work we write $\partial_{\alpha}$ meaning $\frac{\partial}{\partial \theta^\alpha}$. Integration on Grassmann variables is the usual Berezin integration
\begin{align}
&\int d \theta^\alpha \theta_\beta=\delta^\alpha_\beta, & \int d\theta^\alpha 1 =0. 
\end{align}
\\
Equation (\ref{NSusyAlgebra}) reduces to 
\\
\begin{equation} \label{QQN1}
	\{Q_\alpha,Q_\beta\}=2  \gamma^\mu P_\mu.
\end{equation}
\\
The differential representation of $Q_\alpha$ on superspace is 
\\
\begin{equation}
	\mathcal{Q}_\alpha=\partial_\alpha + i \gamma^\mu_{\alpha \beta}\theta^\beta \partial_\mu.
\end{equation}
\\
Equation (\ref{QQN1}) reads
\\
\begin{equation}
	\{\mathcal{Q}_\alpha,\mathcal{Q}_\beta\}=2i\gamma^\mu_{\alpha\beta}\partial_\mu.
\end{equation}
\\
A covariant derivative can be defined as usual requiring that it (anti)commutes with $\mathcal{Q}_\alpha$
\\
\begin{equation} \label{CovDerN1}
	D_\alpha=\partial_\alpha-i\gamma^\mu_{\alpha \beta}\theta^\beta \partial_\mu
\end{equation}
\\
The matter super-multiplet can be assembled into the super-field 
\begin{equation}
	\Phi(x,\theta)=\phi(x)+\theta\psi(x)-\theta \theta F(x)
\end{equation}
\\
where $\phi$ is real scalar field, $\psi$ (real) is Majorana spinor, and $F$ is a non-dynamical scalar field.
\\

\subsection{$\mathcal{N}=2$} \label{appendixA2}
If $\mathcal{N}=2$ there are two spinor generators $Q^I_\alpha$ and $4$ supercharges. The supersymmetry algebra can be obtained by dimensional reduction from the well-known $\mathcal{N}=1$ $4d$ algebra \cite{McKeon} \cite{SEIBERG1} 
\\
\begin{equation} \label{SusyAlgebraReduced}
	\{Q_\alpha,\bar{Q}_\beta\}=2\gamma_{\alpha \beta
	}^\mu P_\mu+2i\varepsilon_{\alpha \beta}Z
\end{equation}
\\
where $Q$ and $\bar{Q}$ are complex and the central charge $Z$ is the momentum in the reduced dimension. Notice that central charges vanishes on massless representations. Alternatively, we can start from the formulation with Majorana fermionic generators as in equation (\ref{NSusyAlgebra}): $Q^1$ and $Q^2$. The change of variables 
\begin{align}
&Q=Q_1+iQ_2, &\bar{Q}=Q_1-iQ_2
\end{align} 
brings equation (\ref{NSusyAlgebra}) into (\ref{SusyAlgebraReduced}) with vanishing $Z$. \\
The differential representation of $Q$ and $\bar{Q}$ is
\\
\begin{align}
	&\mathcal{Q}_\alpha=-\partial_\alpha-\gamma^\mu_{\alpha \beta}\bar{\theta}^{\beta}\partial_{\mu},
	&\bar{\mathcal{Q}}_\alpha=\bar{\partial}_\alpha+\theta^{\beta}\gamma^\mu_{ \beta \alpha}\partial_{\mu}.
\end{align}
\\
Covariant derivatives $ D $ and $\bar{D}$ can be defined as usual requiring that they anti-commute with $Q$ and $\bar{Q}$
\\
\begin{align} \label{DbarD}
	&D_{\alpha}=\partial_{\alpha}+i  \gamma^{\mu}_{\alpha \rho}\bar{\theta}^{\rho} \partial_{\mu} \\\label{DbarD2}
	&\bar{D}_{\alpha}=\bar{\partial}_{\alpha}+i \theta^{\rho}\gamma^{\mu}_{ \rho \alpha} \partial_{\mu}	
\end{align}
\\
and turns out that their (anti)commutation rules are
\\
\begin{align} \label{antiDbarD}
	&\{ D_{\alpha},\bar{D}_{\beta} \} \ = \ 2i \gamma^{\mu}_{\alpha \beta} \partial_{\mu}
	\\
	&\{D_\alpha,D_\beta\} \ = \ 0
	\\
	&\{\bar{D}_\alpha,\bar{D}_\beta\} \ = \ 0
\end{align}
\\
and using the equation (\ref{antiDbarD}) it can be shown that
\begin{equation}
	D^2\bar{D}^2=\DAlambert.
\end{equation}
\\
The chiral and anti-chiral superfields are defined imposing the shortening conditions 
\\
\begin{equation} \label{DQbarDbarQ}
	D_{\alpha} \bar{\Phi} =0
	\ \ \ \ \ \ \ \ \ \ \ \
	\bar{D}_{\alpha}\Phi=0
\end{equation}
\\
Solutions of equations (\ref{DQbarDbarQ}) define chiral and anti-chiral superfields
\\
\begin{equation}
	\Phi=\phi(y)+\theta^{\alpha} \psi_{\alpha}(y)-\theta^2 F(y),
	\ \ \ \ \ \ \ \ \ \
	y^{\mu}=x^{\mu}+i\theta \gamma^{\mu}\bar{\theta},
\end{equation}
\\
\begin{equation}
	\bar{\Phi}=\bar{\phi}(y)+\bar{\theta}^{\alpha} \bar{\psi}_{\alpha}(\bar{y})-\bar{\theta}^2 \bar{F}(\bar{y}),
	\ \ \ \ \ \ \ \ \ \ 
	\bar{y}^{\mu}=x^{\mu}-i\theta \gamma^{\mu}\bar{\theta},
\end{equation}
\\
where  $F$ or $\bar{F}$ are auxiliary fields and not dynamical degrees of freedom.\\
Both the chiral and anti-chiral superfields carry a matter supermultiplet: which has 2 bosonic degrees of freedom (a complex scalar field) and 4 fermionic degrees of freedom (a Dirac spinor).
This means that a (anti)chiral $\mathcal{N}=2$ super-multiplet can be decomposed into two $\mathcal{N}=1$ matter supermultiplets
\\
\begin{equation}
	Chiral_{\mathcal{N}=2} = Matter_{\mathcal{N}=1} \oplus Matter_{\mathcal{N}=1}.
\end{equation}  

\newpage
\section{$\mathcal{N}=1$ computations} \label{appendixB}
In this appendix we show some explicit computations in the $d=3$ superspace. All these results are well know and the interested reader can find more details in \cite{GRIS}.
\subsection{Propagator} \label{appendixB1}
The operator $D^2$ is defined as $D^2=\frac{1}{2}D^{\alpha} D_\alpha$ and can be computed straightforwardly using the explicit form of $D_\alpha$ in equation (\ref{CovDerN1})
\begin{align} \label{DprimeD} \nonumber
	D^{\alpha}D_{\alpha}&=C^{\alpha \beta}D_{\alpha}D_{\beta}= \left(\partial_\alpha-i \gamma^{\mu}_{\alpha \rho}\theta^{\rho}\partial_{\mu} \right) \left(C^{\alpha \beta} \right) \left(\partial_\beta-i \gamma^{\nu}_{\beta \lambda}\theta^{\lambda}\partial_{\nu} \right)
	\\&=\partial^{\alpha} \partial_{\alpha}+i2C^{\alpha \beta}\gamma^{\mu}_{\alpha \rho}\partial_{\mu}\theta
	^{\rho}\partial_{\beta}-C^{\alpha \beta} \gamma_{\alpha \rho}^{\mu}\gamma^{\nu}_{\beta \lambda}\theta^{\rho}\theta^{\lambda}\partial_{\mu}\partial_{\nu}.
\end{align}
We will need also another useful identity involving the $D_{\alpha}$ operator, which is
\begin{equation} \label{D2square}
	\left(D^2\right)^2=\DAlambert.
\end{equation}
Notice that in order to prove (\ref{DprimeD}) it is not necessary to use the explicit form $D^2$. It simply follows from
\begin{equation}
	\{D_{\alpha},D_{\beta}\}\{D^{\alpha},D^{\beta}\}=\left( 2i\gamma^\mu_{\alpha \beta} \partial_\mu \right) C^{\alpha \rho} C^{\beta \sigma} \left( 2i\gamma^\nu_{\rho \sigma} \partial_\nu \right),
\end{equation}
\\
using the algebra of $D_\alpha$ and identities for the trace of gamma matrices\footnote{In particular we used $D_\alpha D_\beta D^\alpha=0$ and $D^2D_\alpha=-D_\alpha D^2$.}.

Thanks to equation (\ref{D2square}) it is trivial to compute the propagator by inverting the quadratic part of the action
\begin{equation}
	\langle \Phi(x_1,\theta_1) \Phi(x_2,\theta_2) \rangle= -\frac{1}{D^2} \delta^{2} ( \theta_1 - \theta_2 )=-\frac{D^2}{\DAlambert} \delta^{2} ( \theta_1 - \theta_2 ).
\end{equation} 
\\
In momentum space the propagator reads
\\
\begin{equation} \label{N1Propagator}
	G_0(p,\theta)=\langle \Phi(-p,\theta_1) \Phi(p,\theta_2) \rangle_0=\frac{D^2\delta^{2} ( \theta_1 - \theta_2 )}{p^2} .
\end{equation} 
\\
However, it is really convenient to recast equation (\ref{N1Propagator}) in a different form. By applying $D^2$ on $\delta^2(\theta_1-\theta_2)$\footnote{For Grassmann variables we have that $\delta^{2}(\theta_1-\theta_2)=(\theta_1-\theta_2)^2$ by definition.} it is easy to show that we can rewrite the numerator of the propagator as
\\
\begin{equation} \label{DsquaredDelta}
	D^2 \delta^2(\theta_1-\theta_2) = 1- \left( \theta_1^\alpha \gamma^\mu_{\alpha \beta} \theta_2^\beta \right)p_\mu-p^2\theta_1^2 \theta_2^2.
\end{equation}
\\
A useful feature of this expression is that it can be exponentiated thanks to the anticommuting nature of Grassmann numbers
\\
\begin{equation} \label{expDsquared}
	D^2 \delta^2(\theta_1-\theta_2)=e^{-\left( \theta_1^\alpha \gamma^\mu_{\alpha \beta} \theta_2^\beta \right)p_\mu}.
\end{equation}
\\
The zeroth order and the first order of the series expansion of (\ref{expDsquared}) obviously reproduce the first two terms in the r.h.s. of (\ref{DsquaredDelta}). Since all terms higher then the third one vanish, it is sufficient to check that second order of the series is $-p^2 \theta_1^2 \theta_2^2$.

\subsection{Computation of $\mathcal{J}(p,q,k,\theta_1,\theta_2)$}\label{appendixB2}
The Grassmann integral $\mathcal{J}(p,q,k,\theta_1,\theta_2)$ has been defined in equations (\ref{DSEN1}, \ref{DSEIRN1}, \ref{JschematicDef}). Using equation we find
\begin{align}
	\mathcal{J}(p,q,k,\theta_1,\theta_2)=\int d^2\theta' d^2 \theta'' &e^{-(\theta_1\gamma^\mu\theta')p_\mu} \left(  e^{-(\theta'\gamma^\mu\theta'')q_{\mu}} e^{-(\theta'\gamma^\mu\theta'')k_{\mu}} e^{-(\theta'\gamma^\mu\theta'')(p-q-k)_{\mu}} \right)\nonumber\\
	& \times e^{-(\theta''\gamma^\mu\theta_2)p_{\mu}}.
\end{align}
\\
Since all exponents are bilinear in $\theta$ they commute among each others and one is free to bring all of them in the same exponential
\\
\begin{equation}
	\mathcal{J}(p,\theta_1,\theta_2)=\int d^2 \theta' d^2 \theta'' e^{-(\theta_1\gamma^\mu\theta')p_\mu}   e^{-(\theta'\gamma^\mu\theta'')p_{\mu}}  e^{-(\theta''\gamma^\mu\theta_2)p_{\mu}}.
\end{equation}
\\
It is remarkable that $k$ and $q$ simplify and $\mathcal{J}(p,q,k,\theta_1,\theta_2)$ is actually a function of $p$, $\theta_1$ and $\theta_2$ only. For this reason,with a slightly abuse of notation, we will start writing $\mathcal{J}(p,\theta_1,\theta_2)$.
This mechanism is really general and will survive in the $\mathcal{N}=2$ case. Each time we write the contribution of a 'melonic' loop in a Feynman diagram, if the superspace part can be written in the exponential form, we can sum the exponents and the integrated momenta cancel. 
\\
Now $\mathcal{J}(p,\theta_1,\theta_2)$ can be integrated straightforwardly and only few terms, those with the correct number of $\theta$s, survive on the integration. It is convenient to use equation (\ref{DsquaredDelta}) and some useful identities listed below
\\
\begin{align} \label{thetathetaC}
	&\theta^\alpha \theta^\beta=C^{\alpha \beta } \theta^2,\\\label{gammagamma}
	&\gamma^\mu \gamma^\nu= \eta^{\mu \nu} \mathbb{I}-\varepsilon^{\mu \nu \rho}\gamma_\rho ,\\\label{BasicRule}
	&\frac{1}{2}\left( \theta_2 \gamma^\mu \theta_1 \right) \left( \theta_2 \gamma^\nu \theta_1 \right)p_\mu p_\nu=-p^2 \theta_2^2 \theta_1^2,\\ \label{BasicRule2}
	&\frac{1}{2}\left( \theta_2 \gamma^\mu \theta_1 \right) \left( \theta_3 \gamma^\nu \theta_2 \right)p_\mu p_\nu=p^2 \theta_2^2 (\theta_1^\alpha \theta_{3\alpha}).
\end{align}
\\
In the end we find 
\\
\begin{equation}
	\mathcal
	{J}(p)=-p^2D^2\delta(\theta_2-\theta_1).
\end{equation}

\subsection{Computation of $\mathcal{J}_{2}(p,q,k,\theta_1,\theta_2)$} \label{appendixB3}
$\mathcal{J}_{2}(p,q,k,\theta_1,\theta_2)$ has been defined in equations (\ref{KV2}, \ref{KBB}, \ref{J2schematic}) and it  writes (schematically\footnote{The reader should look at equation (\ref{KV2}) in order to track back the explicit dependence on momenta.})
\\
\begin{equation}
\mathcal{J}_{2}(p,q,k,\theta_1,\theta_2) \equiv \int d^2\theta' d^2 \theta'' D^2\delta^2(\theta_1-\theta') \left( \left(D^2\delta^2(\theta'-\theta'')\right)^2 \delta(\theta'-\theta'') \right) D^2\delta^2(\theta''-\theta_2).
\end{equation}
\\
The difference with respect to $\mathcal{J}(p,q,k,\theta_1,\theta_2)$ is that one of the factors $D^2\delta^2(\theta'-\theta'')$ has been changed in $\delta^2(\theta'-\theta'')$. Despite the fact that $\delta^2(\theta'-\theta'')$  cannot be exponentiated, the integral is really easy to solve using the fact that
\\
\begin{equation}
	\delta^2(\theta'-\theta'')D^2\delta^2(\theta'-\theta'')=\delta^2(\theta'-\theta'') e^{-(\theta'^\alpha \gamma_{\alpha \beta}^\mu \theta''^\beta)p_\mu}=\delta^2(\theta'-\theta''),
\end{equation}
\\
because gamma matrices with two low indices are symmetric. We see again that $\mathcal{J}_2(p,q,k,\theta_1,\theta_2)$ does not depend on $q$ and $k$, then taking the integral in $d^2 \theta''$ we find
\\
\begin{equation}
	\mathcal{J}_{2}(p,q,k,\theta_1,\theta_2)=\mathcal{J}_{2}(p,\theta_1,\theta_2)=\int d^2\theta_1 D^2\delta^2(\theta'-\theta_1)D^2\delta^2(\theta''-\theta_1).
\end{equation}
\\
Expanding the two exponentials and integrating in $d^2\theta'$ only quadratic terms in $\theta'$ survive and we get the final results
\\
\begin{equation}\label{JBB}
	\mathcal{J}_{2}(p,\theta_1,\theta_2)=-p^2 \delta^2(\theta_2-\theta_1).
\end{equation}

\bibliography{bibliografy}

\providecommand{\href}[2]{#2}\begingroup\raggedright\begin{thebibliography}{10}

\bibitem{ZinnJustin1}
J.~Zinn-Justin, \emph{{Vector models in the large N limit: A Few
  applications}},  in \emph{{11th Taiwan Spring School on Particles and
  Fields}}, 3, 1998 [\href{https://arxiv.org/abs/hep-th/9810198}{{\ttfamily
  hep-th/9810198}}].

\bibitem{Moshe:2003xn}
M.~Moshe and J.~Zinn-Justin, \emph{{Quantum field theory in the large N limit:
  A Review}}, \href{https://doi.org/10.1016/S0370-1573(03)00263-1}{\emph{Phys.
  Rept.} {\bfseries 385} (2003) 69}
  [\href{https://arxiv.org/abs/hep-th/0306133}{{\ttfamily hep-th/0306133}}].

\bibitem{GT1}
G.~'t~Hooft, \emph{A planar diagram theory for strong interaction},
  {\emph{Nuclear Physics} (1974) }.

\bibitem{KlebPop}
I.R.~Klebanov, F.~Popov and G.~Tarnopolsky, \emph{{TASI Lectures on Large $N$
  Tensor Models}}, \href{https://doi.org/10.22323/1.305.0004}{\emph{PoS}
  {\bfseries TASI2017} (2018) 004}
  [\href{https://arxiv.org/abs/1808.09434}{{\ttfamily 1808.09434}}].

\bibitem{Gurau:2019qag}
R.~Gurau, \emph{{Notes on Tensor Models and Tensor Field Theories}},
  \href{https://arxiv.org/abs/1907.03531}{{\ttfamily 1907.03531}}.

\bibitem{BenedettiMelonicCFT}
D.~Benedetti, \emph{{Melonic CFTs}},  in \emph{{19th Hellenic School and
  Workshops on Elementary Particle Physics and Gravity}}, 4, 2020
  [\href{https://arxiv.org/abs/2004.08616}{{\ttfamily 2004.08616}}].

\bibitem{Klebanov:2016xxf}
I.R.~Klebanov and G.~Tarnopolsky, \emph{{Uncolored random tensors, melon
  diagrams, and the Sachdev-Ye-Kitaev models}},
  \href{https://doi.org/10.1103/PhysRevD.95.046004}{\emph{Phys. Rev. D}
  {\bfseries 95} (2017) 046004}
  [\href{https://arxiv.org/abs/1611.08915}{{\ttfamily 1611.08915}}].

\bibitem{Gurau:2009tw}
R.~Gurau, \emph{{Colored Group Field Theory}},
  \href{https://doi.org/10.1007/s00220-011-1226-9}{\emph{Commun. Math. Phys.}
  {\bfseries 304} (2011) 69} [\href{https://arxiv.org/abs/0907.2582}{{\ttfamily
  0907.2582}}].

\bibitem{Witten1}
E.~Witten, \emph{{An SYK-Like Model Without Disorder}},
  \href{https://doi.org/10.1088/1751-8121/ab3752}{\emph{J. Phys. A} {\bfseries
  52} (2019) 474002} [\href{https://arxiv.org/abs/1610.09758}{{\ttfamily
  1610.09758}}].

\bibitem{Bonzom:2012hw}
V.~Bonzom, R.~Gurau and V.~Rivasseau, \emph{{Random tensor models in the large
  N limit: Uncoloring the colored tensor models}},
  \href{https://doi.org/10.1103/PhysRevD.85.084037}{\emph{Phys. Rev. D}
  {\bfseries 85} (2012) 084037}
  [\href{https://arxiv.org/abs/1202.3637}{{\ttfamily 1202.3637}}].

\bibitem{GiombiKleb}
S.~Giombi, I.R.~Klebanov and G.~Tarnopolsky, \emph{{Bosonic tensor models at
  large $N$ and small $\epsilon$}},
  \href{https://doi.org/10.1103/PhysRevD.96.106014}{\emph{Phys. Rev. D}
  {\bfseries 96} (2017) 106014}
  [\href{https://arxiv.org/abs/1707.03866}{{\ttfamily 1707.03866}}].

\bibitem{GurauBen2}
D.~Benedetti, R.~Gurau and S.~Harribey, \emph{{Line of fixed points in a
  bosonic tensor model}},
  \href{https://doi.org/10.1007/JHEP06(2019)053}{\emph{JHEP} {\bfseries 06}
  (2019) 053} [\href{https://arxiv.org/abs/1903.03578}{{\ttfamily
  1903.03578}}].

\bibitem{Benedetti:2019rja}
D.~Benedetti, N.~Delporte, S.~Harribey and R.~Sinha, \emph{{Sextic tensor field
  theories in rank $3$ and $5$}},
  \href{https://doi.org/10.1007/JHEP06(2020)065}{\emph{JHEP} {\bfseries 06}
  (2020) 065} [\href{https://arxiv.org/abs/1912.06641}{{\ttfamily
  1912.06641}}].

\bibitem{Giombi:2018qgp}
S.~Giombi, I.R.~Klebanov, F.~Popov, S.~Prakash and G.~Tarnopolsky,
  \emph{{Prismatic Large $N$ Models for Bosonic Tensors}},
  \href{https://doi.org/10.1103/PhysRevD.98.105005}{\emph{Phys. Rev. D}
  {\bfseries 98} (2018) 105005}
  [\href{https://arxiv.org/abs/1808.04344}{{\ttfamily 1808.04344}}].

\bibitem{MaldacenaStanford}
J.~Maldacena and D.~Stanford, \emph{{Remarks on the Sachdev-Ye-Kitaev model}},
  \href{https://doi.org/10.1103/PhysRevD.94.106002}{\emph{Phys. Rev. D}
  {\bfseries 94} (2016) 106002}
  [\href{https://arxiv.org/abs/1604.07818}{{\ttfamily 1604.07818}}].

\bibitem{ROSsyk}
V.~Rosenhaus, \emph{{An introduction to the SYK model}},
  \href{https://doi.org/10.1088/1751-8121/ab2ce1}{\emph{J. Phys. A} {\bfseries
  52} (2019) 323001} [\href{https://arxiv.org/abs/1807.03334}{{\ttfamily
  1807.03334}}].

\bibitem{GrossRoss}
D.J.~Gross and V.~Rosenhaus, \emph{{A Generalization of Sachdev-Ye-Kitaev}},
  \href{https://doi.org/10.1007/JHEP02(2017)093}{\emph{JHEP} {\bfseries 02}
  (2017) 093} [\href{https://arxiv.org/abs/1610.01569}{{\ttfamily
  1610.01569}}].

\bibitem{POL}
J.~Polchinski and V.~Rosenhaus, \emph{{The Spectrum in the Sachdev-Ye-Kitaev
  Model}}, \href{https://doi.org/10.1007/JHEP04(2016)001}{\emph{JHEP}
  {\bfseries 04} (2016) 001}
  [\href{https://arxiv.org/abs/1601.06768}{{\ttfamily 1601.06768}}].

\bibitem{POP1}
F.K.~Popov, \emph{Supersymmetric tensor model at large $n$ and small
  $\epsilon$}, \href{https://doi.org/10.1103/PhysRevD.101.026020}{\emph{Phys.
  Rev. D} {\bfseries 101} (2020) 026020}
  [\href{https://arxiv.org/abs/1907.02440}{{\ttfamily 1907.02440}}].

\bibitem{Komargodski:2012ek}
Z.~Komargodski and A.~Zhiboedov, \emph{{Convexity and Liberation at Large
  Spin}}, \href{https://doi.org/10.1007/JHEP11(2013)140}{\emph{JHEP} {\bfseries
  11} (2013) 140} [\href{https://arxiv.org/abs/1212.4103}{{\ttfamily
  1212.4103}}].

\bibitem{Fitzpatrick:2012yx}
A.~Fitzpatrick, J.~Kaplan, D.~Poland and D.~Simmons-Duffin, \emph{{The Analytic
  Bootstrap and AdS Superhorizon Locality}},
  \href{https://doi.org/10.1007/JHEP12(2013)004}{\emph{JHEP} {\bfseries 12}
  (2013) 004} [\href{https://arxiv.org/abs/1212.3616}{{\ttfamily 1212.3616}}].

\bibitem{Alday:2016njk}
L.F.~Alday, \emph{{Large Spin Perturbation Theory for Conformal Field
  Theories}}, \href{https://doi.org/10.1103/PhysRevLett.119.111601}{\emph{Phys.
  Rev. Lett.} {\bfseries 119} (2017) 111601}
  [\href{https://arxiv.org/abs/1611.01500}{{\ttfamily 1611.01500}}].

\bibitem{Alday:2016jfr}
L.F.~Alday, \emph{{Solving CFTs with Weakly Broken Higher Spin Symmetry}},
  \href{https://doi.org/10.1007/JHEP10(2017)161}{\emph{JHEP} {\bfseries 10}
  (2017) 161} [\href{https://arxiv.org/abs/1612.00696}{{\ttfamily
  1612.00696}}].

\bibitem{Caron-Huot:2017vep}
S.~Caron-Huot, \emph{{Analyticity in Spin in Conformal Theories}},
  \href{https://doi.org/10.1007/JHEP09(2017)078}{\emph{JHEP} {\bfseries 09}
  (2017) 078} [\href{https://arxiv.org/abs/1703.00278}{{\ttfamily
  1703.00278}}].

\bibitem{PARK1}
J.-H.~Park, \emph{Superconformal symmetry in three dimensions},
  \href{https://doi.org/10.1063/1.1290056}{\emph{Journal of Mathematical
  Physics} {\bfseries 41} (2000) 7129}.

\bibitem{ATA1}
A.~Atanasov, A.~Hillman and D.~Poland, \emph{{Bootstrapping the Minimal 3D
  SCFT}}, \href{https://doi.org/10.1007/JHEP11(2018)140}{\emph{JHEP} {\bfseries
  11} (2018) 140} [\href{https://arxiv.org/abs/1807.05702}{{\ttfamily
  1807.05702}}].

\bibitem{NIZAMI}
A.A.~Nizami, T.~Sharma and V.~Umesh, \emph{{Superspace formulation and
  correlation functions of 3d superconformal field theories}},
  \href{https://doi.org/10.1007/JHEP07(2014)022}{\emph{JHEP} {\bfseries 07}
  (2014) 022} [\href{https://arxiv.org/abs/1308.4778}{{\ttfamily 1308.4778}}].

\bibitem{Bobev}
N.~Bobev, S.~El-Showk, D.~Mazac and M.F.~Paulos, \emph{{Bootstrapping SCFTs
  with Four Supercharges}},
  \href{https://doi.org/10.1007/JHEP08(2015)142}{\emph{JHEP} {\bfseries 08}
  (2015) 142} [\href{https://arxiv.org/abs/1503.02081}{{\ttfamily
  1503.02081}}].

\bibitem{CordovaInt}
C.~Cordova, T.T.~Dumitrescu and K.~Intriligator, \emph{{Multiplets of
  Superconformal Symmetry in Diverse Dimensions}},
  \href{https://doi.org/10.1007/JHEP03(2019)163}{\emph{JHEP} {\bfseries 03}
  (2019) 163} [\href{https://arxiv.org/abs/1612.00809}{{\ttfamily
  1612.00809}}].

\bibitem{Gracey}
J.~Gracey, I.~Jack, C.~Poole and Y.~Schröder, \emph{{a-function for $N=$ 2
  supersymmetric gauge theories in three dimensions}},
  \href{https://doi.org/10.1103/PhysRevD.95.025005}{\emph{Phys. Rev. D}
  {\bfseries 95} (2017) 025005}
  [\href{https://arxiv.org/abs/1609.06458}{{\ttfamily 1609.06458}}].

\bibitem{toappear}
J.~Henriksson, D.~Lettera and A.~Vichi, \emph{{To appear}}, .

\bibitem{Alday:2019clp}
L.F.~Alday, J.~Henriksson and M.~van Loon, \emph{{An alternative to diagrams
  for the critical O(N) model: dimensions and structure constants to order
  1/N$^{2}$}}, \href{https://doi.org/10.1007/JHEP01(2020)063}{\emph{JHEP}
  {\bfseries 01} (2020) 063}
  [\href{https://arxiv.org/abs/1907.02445}{{\ttfamily 1907.02445}}].

\bibitem{Henriksson:2020fqi}
J.~Henriksson, S.R.~Kousvos and A.~Stergiou, \emph{{Analytic and Numerical
  Bootstrap of CFTs with $O(m)\times O(n)$ Global Symmetry in 3D}},
  \href{https://doi.org/10.21468/SciPostPhys.9.3.035}{\emph{SciPost Phys.}
  {\bfseries 9} (2020) 035} [\href{https://arxiv.org/abs/2004.14388}{{\ttfamily
  2004.14388}}].

\bibitem{Hogervorst:2015akt}
M.~Hogervorst, S.~Rychkov and B.C.~van Rees, \emph{{Unitarity violation at the
  Wilson-Fisher fixed point in 4-$\epsilon$ dimensions}},
  \href{https://doi.org/10.1103/PhysRevD.93.125025}{\emph{Phys. Rev. D}
  {\bfseries 93} (2016) 125025}
  [\href{https://arxiv.org/abs/1512.00013}{{\ttfamily 1512.00013}}].

\bibitem{Cardy:2013rqg}
J.~Cardy, \emph{{Logarithmic conformal field theories as limits of ordinary
  CFTs and some physical applications}},
  \href{https://doi.org/10.1088/1751-8113/46/49/494001}{\emph{J. Phys. A}
  {\bfseries 46} (2013) 494001}
  [\href{https://arxiv.org/abs/1302.4279}{{\ttfamily 1302.4279}}].

\bibitem{Hogervorst:2016itc}
M.~Hogervorst, M.~Paulos and A.~Vichi, \emph{{The ABC (in any D) of Logarithmic
  CFT}}, \href{https://doi.org/10.1007/JHEP10(2017)201}{\emph{JHEP} {\bfseries
  10} (2017) 201} [\href{https://arxiv.org/abs/1605.03959}{{\ttfamily
  1605.03959}}].

\bibitem{Chang:2019yug}
C.-M.~Chang, S.~Colin-Ellerin and M.~Rangamani, \emph{{Supersymmetric
  Landau-Ginzburg Tensor Models}},
  \href{https://doi.org/10.1007/JHEP11(2019)007}{\emph{JHEP} {\bfseries 11}
  (2019) 007} [\href{https://arxiv.org/abs/1906.02163}{{\ttfamily
  1906.02163}}].

\bibitem{GRIS}
S.J.~Gates, M.T.~Grisaru, M.~Rocek and W.~Siegel, \emph{{Superspace Or One
  Thousand and One Lessons in Supersymmetry}}, vol.~58 of \emph{Frontiers in
  Physics} (1983), [\href{https://arxiv.org/abs/hep-th/0108200}{{\ttfamily
  hep-th/0108200}}].

\bibitem{McKeon}
D.~McKeon and T.~Sherry, \emph{{Supersymmetry in three-dimensions}},
  \href{https://arxiv.org/abs/hep-th/0108074}{{\ttfamily hep-th/0108074}}.

\bibitem{SEIBERG1}
O.~Aharony, A.~Hanany, K.A.~Intriligator, N.~Seiberg and M.~Strassler,
  \emph{{Aspects of N=2 supersymmetric gauge theories in three-dimensions}},
  \href{https://doi.org/10.1016/S0550-3213(97)00323-4}{\emph{Nucl. Phys. B}
  {\bfseries 499} (1997) 67}
  [\href{https://arxiv.org/abs/hep-th/9703110}{{\ttfamily hep-th/9703110}}].

\end{thebibliography}\endgroup
\bibliographystyle{JHEP}
\end{document}